\documentclass[12pt,onecolumn,draftclsnofoot]{IEEEtran} 
\usepackage{epsfig} 
\usepackage{graphicx}

\long\def\omitit#1{}
\newcommand{\ie}{{\it i.e.,$\ $}}

\begin{document}

\title{The Low Latency Fault Tolerance System}

\author{Wenbing Zhao,
P. M. Melliar-Smith,
L. E. Moser
\IEEEcompsocitemizethanks{
\IEEEcompsocthanksitem W. Zhao is with the
Department of Electrical and Computer Engineering,
Cleveland State University, Cleveland, OH 44115.\protect\\
\IEEEcompsocthanksitem P. M. Melliar-Smith and L. E. Moser are with the
Department of Electrical and Computer Engineering,
University of California, Santa Barbara, CA 93106.} 
\thanks{}
}

\vspace*{-0.1in}
\begin{abstract} 
\vspace*{-0.05in} 
The Low Latency Fault Tolerance (LLFT) system provides fault tolerance
for distributed applications, using the leader-follower replication
technique.  The LLFT system provides application-transparent
replication, with strong replica consistency, for applications that
involve multiple interacting processes or threads. The LLFT system
comprises a Low Latency Messaging Protocol, a Leader-Determined
Membership Protocol, and a Virtual Determinizer Framework. The Low
Latency Messaging Protocol provides reliable, totally ordered message
delivery by employing a direct group-to-group multicast, where the
message ordering is determined by the primary replica in the group.  The
Leader-Determined Membership Protocol provides reconfiguration and
recovery when a replica becomes faulty and when a replica joins or
leaves a group, where the membership of the group is determined by the primary
replica. The Virtual Determinizer Framework captures the ordering
information at the primary replica and enforces the same
ordering at the backup replicas for major sources of non-determinism,
including multi-threading, time-related operations and socket
communication.  The LLFT system achieves low latency message delivery
during normal operation and low latency reconfiguration and recovery
when a fault occurs.
\end{abstract}

\vspace*{-0.1in}
\begin{keywords}
\vspace*{-0.05in}
distributed applications; distributed systems; fault tolerance; 
performance; reliability, availability, serviceability 
\end{keywords}

\maketitle
\IEEEpeerreviewmaketitle

\vspace*{-0.1in}
\section{Introduction}
\vspace*{-0.05in}
The Low Latency Fault Tolerance (LLFT) system provides fault tolerance
for distributed applications, using the leader-follower replication
technique.  LLFT provides application-transparent replication, with
strong replica consistency, for applications that involve multiple
interacting processes or threads. LLFT supports client-server
applications where both client and server processes are replicated,
and three-tier applications where middle-tier processes are
replicated.  LLFT provides fault tolerance for distributed
applications deployed over a local-area network, as in a data center,
rather than over a wide-area network, such as the Internet.

\begin{figure*}[t]
\begin{center} 
\leavevmode
\epsfxsize=5.6in 
\epsfbox{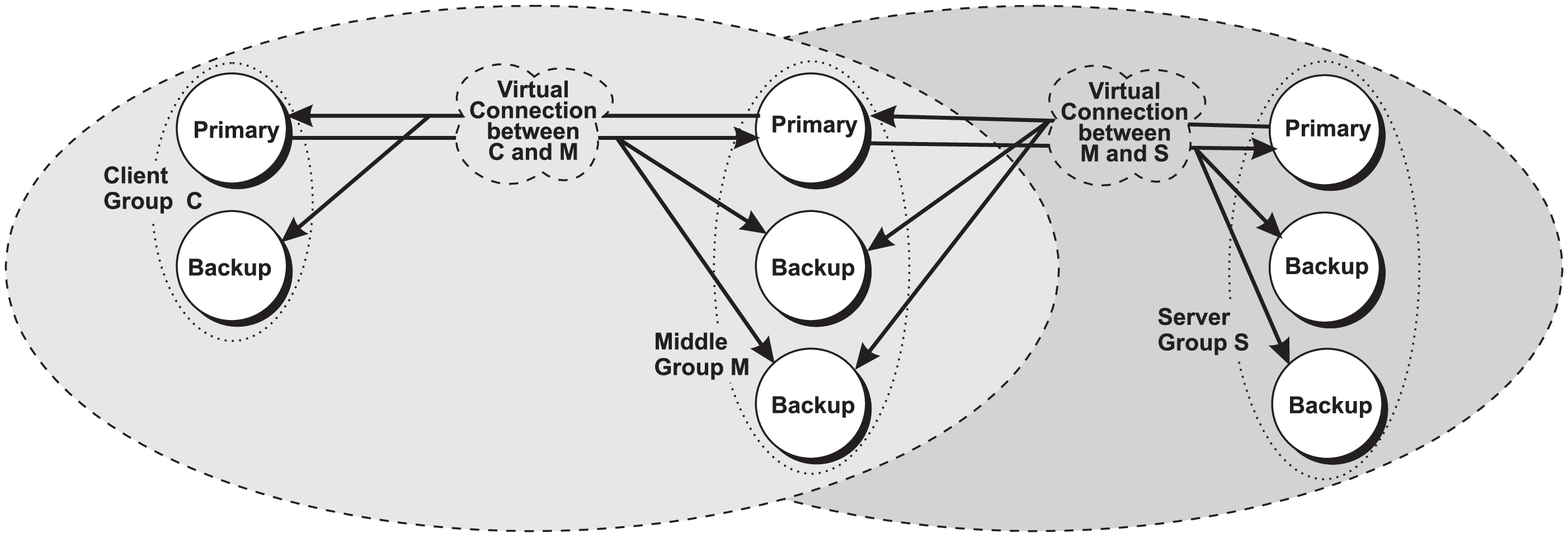}
\vspace*{-0.1in} 
\caption{Process groups interacting over virtual connections.}
\label{conn}
\vspace{-0.2in} 
\end{center}
\end{figure*}

With LLFT, the processes of the application are replicated, and the
replicas of a process form a process group.  Within a process group,
one replica is designated as the primary replica, and the other
replicas are the backup replicas.  The primary in a process group
multicasts messages to a destination process group over a virtual
connection, as shown in Figure~\ref{conn}.  The primary in the
destination process group orders the messages, performs the
operations, and produces ordering information for non-deterministic
operations, which it supplies to the backups in the destination group.

The LLFT system provides fault tolerance for the distributed
applications, with the following properties.

{\bf Strong Replica Consistency.}  The LLFT system replicates the
processes of an application, and maintains strong replica consistency
within a primary component.  The application continues to run
without loss of processing or messages, and without disruption to its
state. If a fault occurs, LLFT provides reconfiguration and recovery
while maintaining virtual synchrony \cite{BR:ISIS,Moser}, including transfer of
state from an existing replica to a new replica and synchronization of
the operation of the new replica with the existing replicas.  To
maintain strong replica consistency within a primary component,
LLFT sanitizes (masks) non-deterministic operations, including
multi-threading, time-related operations and socket communication.

{\bf Low Latency.}  The LLFT system achieves low latency message
delivery during normal operation, and low latency reconfiguration and
recovery when a fault occurs. That is, it provides fault tolerance to
the applications with minimal overhead in the response times seen by
the clients.  LLFT achieves low latency by design, in that the primary
makes the decisions on the order in which operations are performed and
the ordering information is reflected to its backups.  Moreover, the
replicated applications interact with each other directly, without an
intermediate daemon process and without additional context switches.

{\bf Transparency and Ease of Use.}  The LLFT system provides fault
tolerance that is transparent to the application.  The application is
unaware that it is replicated, and is unaware of faults.  Applications
programmed using TCP socket APIs, or middleware such as Java RMI, can
be replicated without modifications to the applications.  The
application programs require no extra code for fault tolerance, and
the application programmers require no special skills in fault
tolerance programming.  The application program is identical to that
of a non-fault-tolerant unreplicated applications.

The novel contributions of this work lie in the design of the
components of the LLFT system.

{\bf Low Latency Messaging Protocol.}  The Low Latency Messaging
Protocol provides reliable, totally ordered message delivery by
communicating message ordering information from the primary replica to
the backup replicas in a group.  It ensures that, in the event of a
fault, a backup has, or can obtain, the messages and the order
information that it needs to reproduce the actions of the primary.
The replicated applications interact with each other directly, via a
group-to-group multicast.

{\bf Leader-Determined Membership Protocol.} The Leader-Determined
Membership Protocol ensures that the members of a process group have a
consistent view of the membership set and of the primary replica in
the group.  It effects a membership change and a consistent view more
quickly than other membership protocols, by selecting a new primary
deterministically, based on the precedences and ranks (defined below)
of the backups in the group and by avoiding the need for a multi-round
consensus algorithm.

{\bf Virtual Determinizer Framework.} The Virtual Determinizer
Framework renders the replicas of an application virtually
deterministic by recording the order and results of each
non-deterministic operation at the primary, and by guaranteeing that
the backups obtain the same results in the same order as the primary.  The
Virtual Determinizer Framework has been instantiated for major sources
of non-determinism, including multi-threading, clock-related
operations and socket communication.

\vspace*{-0.1in}
\section{Basic Concepts}
\label{BasicConcepts}
\vspace*{-0.1in}
\subsection{System Model}
\vspace*{-0.05in}
LLFT operates in an asynchronous distributed system that comprises one
or more applications running on multiple processors and communicating
over a local-area network, such as an Ethernet. Clients that run
outside the local-area network are supported via a gateway.
An application consists of one or more processes, possibly
multi-threaded with shared data, that interact with each other and
also with file systems and database systems. 

A process that is non-faulty completes a computation, but there is no
bound on the time required to complete the computation.  The processes
communicate via messages using an unreliable, unordered message
delivery service, such as UDP multicast, with no bound on the time
required to communicate a message.  With asychronous messaging, no
protocol can guarantee reliable message delivery within a time bound.
Thus, we adopt the (theoretical) assumption of {\it eventual reliable
  communication}, {\it i.e.}, if a message is transmitted repeatedly,
it is eventually received, as do other researchers \cite{Guerraoui}.

\vspace*{-0.1in}
\subsection{Fault Model}
\vspace*{-0.05in}
The LLFT system replicates application processes to protect the
application against various types of faults, in particular:
\begin{itemize}
 \item {\it\bf Crash fault} - A process does not produce
any further results.  
 \item {\it\bf Timing fault} - A process does not produce
a result within a given time constraint.
\end{itemize}
LLFT does not handle Byzantine faults.  LLFT allows processes to
recover but, when a process recovers, it is regarded as a new process
with a new identity ({\tt birthId}).  LLFT also handles communication
network faults, including message loss, selective communication loss,
and partitioning faults.  Healing of partitioning faults is guaranteed
by the eventual reliable communication assumption.

To achieve liveness and termination of the algorithms, LLFT uses
unreliable fault detectors based on timeouts.  The fault detectors are
necessarily unreliable, and the timeouts are a measure of how
unreliable the fault detectors are.  Crash faults are detected as
timing faults by the LLFT fault detectors.

LLFT uses a leader-follower algorithm to establish the total order of
messages, to render operations virtually deterministic, and to
establish a consistent group membership.  It does not use a consensus
algorithm based on unreliable fault detectors \cite{Chandra:UnrelFD}
to circumvent the impossibility result \cite{FLP:IMPOS}.  Moreover,
LLFT does not assume that a majority of the processes in a group is
non-faulty, as do other works such as
\cite{Chandra:UnrelFD,Paxos,Liskov}.  Rather, LLFT adopts the
(theoretical) assumption of {\it sufficient replication}, {\it i.e.},
in each primary view, there exists at least one replica that does not
become faulty.  The price of this relaxation is that LLFT must be able
to detect and heal partitions of the group membership, which it does
using a precedence mechanism (described below).  The risk of multiple
concurrent memberships, and thus the need to detect and heal
partitions, can be avoided only in systems in which a majority vote is
used not only for membership, but also for every value that is
communicated, as is done in aircraft flight control systems
\cite{SIFT}.

LLFT ensures that only one component of a partition, which we refer to
as the {\it primary component}, survives in the infinite sequence of
consecutive primary views of a group.  Within that primary component,
LLFT maintains {\it virtual synchrony} \cite{BR:ISIS,Moser} which
means that, if the primary fails, the new primary must advance to the
state of the old primary, in particular the state known to the remote
groups of its connections, before it failed.

\vspace*{-0.1in}
\subsection{Process Groups}
\vspace*{-0.05in}
The replicas of a process form a {\it process group (virtual
  process)}.  Each process group has a unique identifier ({\it group
  id}), which is supplied by the user.  We equate a process group with its
group id.  This group id is mapped by LLFT to a virtual port on which
the group sends and receives messages, as discussed below.

Each process group has a {\it group membership} that consists of the
replicas of the process. Typically, different members of a process
group run on different processors.  One of the replicas in the process
group is the {\it primary replica}, and the other members are the {\it
  backup replicas}.  Each membership change that introduces a new
primary replica constitutes a new {\it primary view}, which has a {\it
  primary view number}.  Each member of a process group must know the
primary replica in its group.  There is no need for the members of a
sending process group to know which member of a destination process
group is the primary replica.

\vspace*{-0.1in}
\subsection{Virtual Connections}
\vspace*{-0.05in}
The LLFT system introduces the novel, elegant idea of a virtual
connection, which is a natural extension of the point-to-point
connection of TCP.

A {\it virtual connection} is a connection between two endpoints,
where each endpoint is a process group, over which messages are
communicated between the two group endpoints.  A virtual connection is
a full-duplex, many-to-many communication channel between the two
endpoints.  A sender uses UDP multicast to send messages to a
destination group over the virtual connection.

A {\it virtual port (group id)} identifies the source (destination)
group from (to) which the messages are sent (delivered) over the
virtual connection.  All members of a group listen on the same virtual
port, and members of different groups listen on different virtual
ports.  The groups need to know the virtual ports (group ids) of the
groups with which they are communicating, just as with TCP.

Typically, a process group is an endpoint of more than one virtual
connection, as shown in Figure \ref{conn}, where there are multiple
process groups, representing multiple applications running on
different processors and interacting over the network, but there might
be only two process groups and one virtual connection.

\vspace*{-0.1in}
\subsection{Replication Types}
\vspace*{-0.05in}
The LLFT system supports two types of leader-follower replication, namely:
\begin{itemize}
  \item {\it\bf Semi-active replication} - The primary
    orders the messages it receives, performs the 
    operations, and provides ordering information for
    non-deterministic operations to the backups.  
    A backup receives and logs incoming messages, performs the
    operations according to the ordering information supplied by the
    primary, and logs outgoing messages, but does not send
    outgoing messages.
  \item {\it\bf Semi-passive replication} - The primary
    orders the messages it receives, performs the 
    operations, and provides ordering information for
    non-deterministic operations to the backups.  
    In addition, the primary communicates state updates to the
    backups, as well as file and database updates.  A backup receives
    and logs incoming messages, and updates its state, but does not
    perform the operations and does not produce outgoing
    messages.
\end{itemize}
Semi-passive replication uses fewer processing resources than
semi-active replication, but it incurs greater latency for
reconfiguration and recovery, if the primary fails.

To maintain strong replica consistency within a primary component, it
is necessary to sanitize (mask) non-deterministic operations not only
for semi-active replication but also for semi-passive replication.
For example, consider requests from two clients that are processed
concurrently using semi-passive replication.  Processing the request
from the first client updates a data item.  Processing the request
from the second client then updates the same data item.  The request
from the second client completes, and the primary sends its update to
the backups and its reply to the second client.  The primary then
fails before it can send its update to the backups and its reply to
the first client.  The processing of the request from the first client
is repeated at the new primary. However, the update to the data item
has already been performed and recorded at the backup, before it
became the new primary, when the reply was sent to the second client.
The update must not be repeated if the correct result is to be
obtained.

\vspace*{-0.1in}
\subsection{Correctness Properties}
\vspace*{-0.05in}
The safety and liveness properties for LLFT, based on the above model,
are stated below.  Traditionally, safety and liveness properties are
strictly separated.  However, in a system that might incur a
communication partitioning fault with subsequent recovery from that
partitioning, safety is necessarily dependent on liveness.  While the
system is partitioned, even complete knowledge of all processes does
not suffice to determine which of the two or more competing branches
is a transient side branch that will be terminated when the partition
is healed.  Thus, the safety properties for LLFT are defined in terms
of an infinite sequence of consecutive primary views, as assured by
the liveness properties.  The proofs of correctness can be found in
the Appendix.  A discussion of the assumptions of the model, relative
to these properties, is included below.

\subsubsection{{\bf Safety Properties}}
For each process group:
\begin{itemize}
\item There exists at most one infinite sequence of consecutive
  primary views for the group.  Each of those primary views has a 
  unique primary view number and a single primary replica.

\item There exists at most one infinite sequence of operations
  in an infinite sequence of consecutive primary views for the group.

\item For semi-active replication, the sequence of operations of a
  replica, in a membership of the infinite sequence of consecutive
  primary views, is a consecutive subsequence of the infinite sequence
  of operations for the group.

\item For semi-passive replication, the sequence of states of a
  replica, in a membership of the infinite sequence of consecutive
  primary views, is a consecutive subsequence of the infinite sequence
  of states for the group.
\end{itemize}

\subsubsection{{\bf Liveness Properties}}
For each process group:
\begin{itemize}
\item There exists at least one infinite sequence of consecutive
  primary views with consecutive primary view numbers for the group.
\item There exists at least one infinite sequence of operations in
  each infinite sequence of consecutive primary views for the group.
\end{itemize}

LLFT imposes implicit bounds on the computation time and the
communication time in the form of tunable timeout parameters of the
fault detectors. Due to the asynchrony of the system, those bounds
might be violated, which might lead to a replica being mistakenly
removed from the membership.

With the assumption of eventual reliable communication ({\it i.e.}, if
a message is transmitted repeatedly, it is eventually received), a
replica that is mistakenly removed from the membership eventually
receives a {\tt PrimaryChange} or {\tt RemoveBackup} message that
informs it that it has been removed.  The replica then applies for
readmission to the membership using the {\tt ProposeBackup} message.
Without the assumption of eventual reliable communication, a
mistakenly removed replica might not receive those messages and, thus,
not apply for readmission.

The tuning of the timeout parameters for the fault detectors, relative
to the distribution of the times for computation and communication,
can be viewed as a probabilistic optimization problem.  The latency to
determine a new primary and a new membership after detection of a
genuine fault (true positive) is matched against the latency caused by
detection of a mistaken fault (false positive), to minimize the
overall latency of the system.

With the assumption of sufficient replication ({\it i.e.}, each group
contains enough replicas such that in each primary view there exists a
replica that does not become faulty), the sequence of operations of a
group is infinite.  Without the assumption of sufficient replication,
the sequence of operations of a group might be finite.

\vspace*{-0.1in}
\section{Low Latency Messaging Protocol}
\vspace*{-0.05in}
The LLFT Messaging Protocol converts the unreliable, unordered
message delivery service of UDP multicast into a reliable, totally ordered
message delivery service between two group endpoints, just as TCP
converts the unreliable message delivery service of IP unicast into a
reliable, totally ordered message delivery service between two
individual endpoints.

The Messaging Protocol provides the following services for
the application messages:
\begin{itemize}
 \item {\it\bf Reliable delivery} - All of the members in a group
receive each message that is multicast to the group on a connection.
 \item {\it\bf Total ordering} - All of the members in a group deliver
   the messages to the application in the same sequence.
 \item {\it\bf Buffer management} - When a message no longer needs to
be retransmitted (because all of the intended destinations have
received the message), the source and the destinations remove the
message from their buffers.

\end{itemize}
The Messaging Protocol provides reliable, totally ordered message
delivery, while maintaining virtual synchrony \cite{BR:ISIS,Moser} in the
event of a fault.  It incorporates flow control mechanisms similar to
those used by TCP. Flow control is needed to ensure that processing in
the primary receiver, and in the backups, can keep up with the primary
sender, and that buffer space does not become exhausted.

\renewcommand{\tabcolsep}{0.05in}
\renewcommand{\baselinestretch}{0.8}
\begin{figure}\hbox{
\begin{tabular}{ll}
&\parbox[t]{2.5in}{\footnotesize\bf \hspace*{0.5in}Message Types}\\[0.01in]
{\footnotesize\tt\bf Request}&
\parbox[t]{2.5in}{\footnotesize A message that carries application
   payload and that is sent by a primary client.}\\[0.01in]
{\footnotesize\tt\bf Reply}&
\parbox[t]{2.5in}{\footnotesize A message that carries application
   payload and that is sent by a primary server.}\\[0.01in]
{\footnotesize\tt\bf FirstAck}&
\parbox[t]{2.5in}{\footnotesize A control message that is sent by a primary 
   to acknowledge the receipt of a {\tt Request} or {\tt Reply} message.}\\[0.01in]
{\footnotesize\tt\bf SecondAck}&
\parbox[t]{2.5in}{\footnotesize A control message sent by a backup
   to acknowledge the receipt of a {\tt FirstAck} message.}\\[0.01in]
{\footnotesize\tt\bf Nack}&
\parbox[t]{2.5in}{\footnotesize A control message sent by a backup to its
   primary, or by a primary to the primary that originated a
   missing {\tt Request} or {\tt Reply} message, which then retransmits 
   the message.}\\[0.01in]
{\footnotesize\tt\bf Heartbeat}&
\parbox[t]{2.5in}{\footnotesize A control message sent by the primary
   and the backups to facilitate fault detection and also to share 
   timestamp watermark information, which is used for buffer management.}\\[0.01in]
{\footnotesize\tt\bf KeepAlive}&
\parbox[t]{2.5in}{\footnotesize A control message sent by 
   interacting groups over an inter-group connection, to indicate
   the liveliness of the connection, after the connection has been
   idle more than a predetermined amount of time.}\\ \\ \\ \\ \\
\end{tabular}
\hspace*{0.0in}
\begin{tabular}{ll}
&\parbox[t]{2.8in}{\footnotesize\bf \hspace*{0.5in}Message Header Fields}\\[0.01in]
{\footnotesize\tt\bf messageType}&
\parbox[t]{2.8in}{\footnotesize The type of message ({\tt Request}, {\tt Reply}, {\tt FirstAck}, {\tt SecondAck}, {\tt Nack}, {\tt Heartbeat}, {\tt KeepAlive}).}\\[0.01in]
{\footnotesize\tt\bf sourceGroupId}&
\parbox[t]{2.8in}{\footnotesize The identifier of the source group of the message.}\\[0.01in]
{\footnotesize\tt\bf destGroupId}&
\parbox[t]{2.8in}{\footnotesize The identifier of the destination group of the message.}\\[0.01in]
{\footnotesize\tt\bf connSeqNum}&
\parbox[t]{2.8in}{\footnotesize A connection sequence number used to
   identify the connection on which the message is sent.}\\[0.01in]
{\footnotesize\tt\bf primaryViewNum}&
\parbox[t]{2.8in}{\footnotesize The primary view number, a sequence
number that represents the number of membership changes that involve a
change in the primary.}\\[0.01in]
{\footnotesize\tt\bf precedence}&
\parbox[t]{2.8in}{\footnotesize The precedence of the primary.}\\[0.01in]
{\footnotesize\tt\bf msgSeqNum}&
\parbox[t]{2.8in}{\footnotesize The message sequence number, which is
  non-zero if and only if the message is a {\tt Request} or {\tt Reply} 
message multicast by the primary.}\\[0.01in]
{\footnotesize\tt\bf ackViewNum}&
\parbox[t]{2.8in}{\footnotesize The primary view number of the message with
the message sequence number in the {\tt ack} field.}\\[0.01in]
{\footnotesize\tt\bf ack}&
\parbox[t]{2.8in}{\footnotesize A message sequence number, which 
   is non-zero if and only if the message is a {\tt Request} or
   {\tt Reply} message, and the primary has received all
   messages on the connection with sequence numbers less than or equal
   to this sequence number.}\\[0.01in]
{\footnotesize\tt\bf back}&
\parbox[t]{2.8in}{\footnotesize A timestamp watermark used for buffer
   management to indicate that all members of a group have
   received all messages with timestamps less than this 
   watermark.}\\[0.01in]
{\footnotesize\tt\bf timestamp}&
\parbox[t]{2.8in}{\footnotesize A timestamp derived from a Lamport logical
   clock at the source of the message.}\\ \\ \\
\end{tabular}}
\vspace*{-0.2in}
\caption{The message types and the message header fields used by the Messaging 
Protocol.}
\label{MessageTypesMessageHeaders}
\end{figure}
\renewcommand{\baselinestretch}{1.5}

\vspace*{-0.1in}
\subsection{Data Structures}
\vspace*{-0.05in}
\subsubsection{{\bf Message Types}}
\begin{figure}[t]
\begin{center} 
\leavevmode
\epsfxsize=3.3in 
\epsfbox{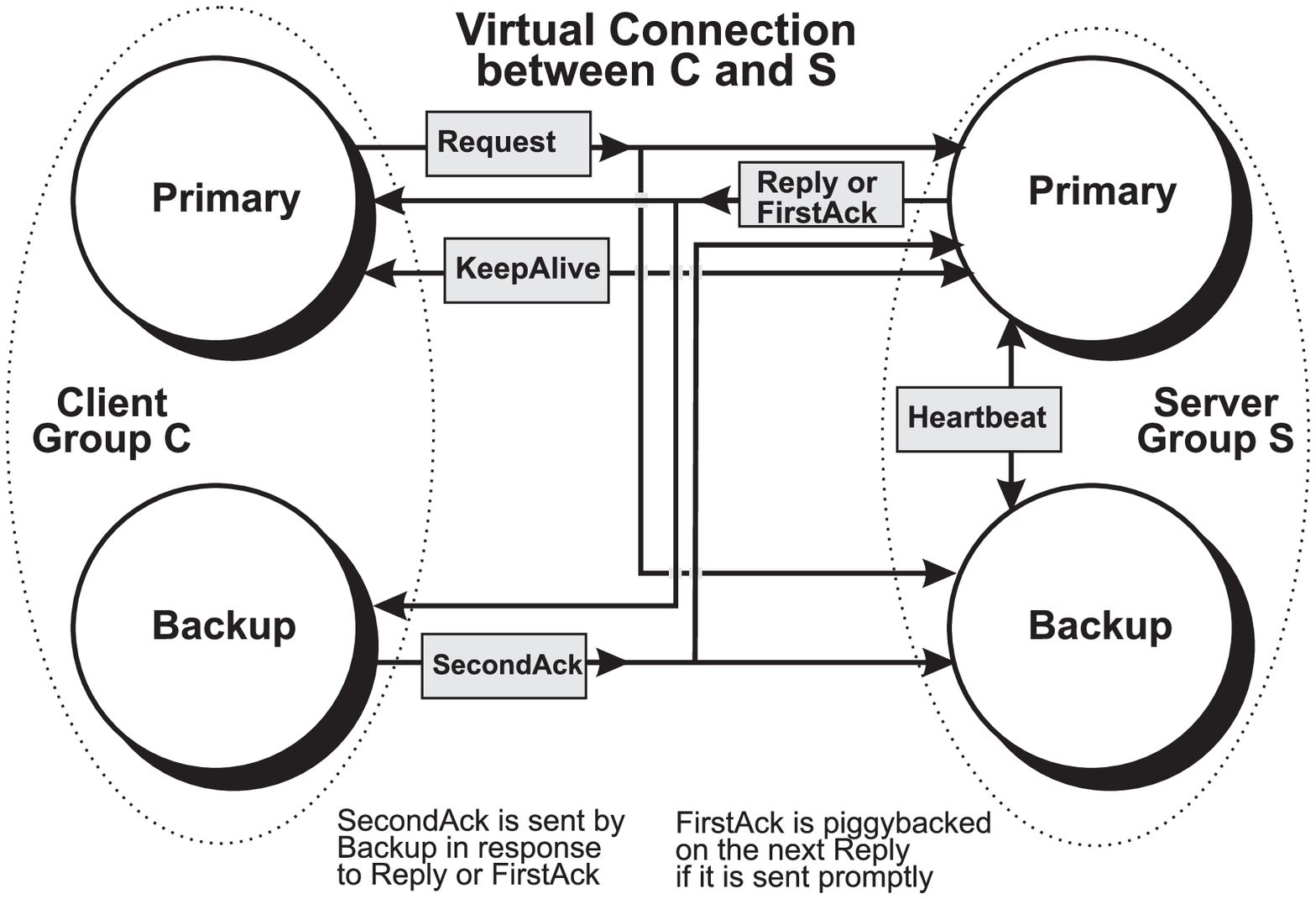}
\vspace{-0.15in}
\caption{Message exchange between client and server groups.}
\label{reliable-messaging}
\vspace{-0.15in}
\end{center}
\end{figure}

The types of messages used by the Messaging Protocol are shown on the
left of Figure~\ref{MessageTypesMessageHeaders} and are illustrated in
Figure \ref{reliable-messaging}. A {\tt Request} or {\tt Reply}
message is not necessarily a synchronous blocking request or reply, as
is commonly used in client/server communication; a {\tt Request} or
{\tt Reply} message can be an asynchronous one-way message.  A
retransmitted {\tt Request} or {\tt Reply} message uses the same
message type as the original message.

\subsubsection{{\bf Message Header Fields}}
The fields of a message header are shown on the right of
Figure~\ref{MessageTypesMessageHeaders}.  The quadruple ({\tt
  sourceGroupId}, {\tt destGroupId}, {\tt connSeqNum}, {\tt role})
uniquely identifies a connection, where {\tt role} is client or
server.

A message with a non-zero message sequence number {\tt msgSeqNum},
{\it i.e.}, a {\tt Request} or {\tt Reply} message multicast by the
primary is inserted into the sent list at the sender and the received
list at the destination.  An {\tt ack} acknowledges not only the
acknowledged message but also all prior messages from the remote
primary of the connection, and allows more rapid retransmission and
delivery of missing messages.

In a message multicast by the primary on a connection, {\tt back}
contains the primary's {\tt myGroup} {\tt Watermark}, \ie the minimum
timestamp watermark of the group, which is the minimum of the
primary's own {\tt myTime} {\tt stampWatermark} and all of the
backups' {\tt myTimestampWatermark}s. In a control message sent by a
backup to its primary, {\tt back} contains the backup's {\tt
  myTimestampWatermark}, \ie the minimum timestamp of messages that
the backup received on all of its connections.  The {\tt timestamp},
which drives the {\tt back}s, is used for buffer management and not
for message ordering.

\subsubsection{{\bf Variables}}
To achieve reliable, totally ordered message delivery, for each
connection, the Messaging Protocol uses the variables shown on the
left of Figure~\ref{Variables}.  
The message sequence number is used by a member of the destination
group to ensure that it has received all messages from the sending
group on the connection.
When a member acknowledges the message with message sequence number {\tt
  receivedUpToMsn}, it indicates that it has received all messages,
sent on the connection, with message sequence numbers less than or equal to
this sequence number.

For buffer management, the Messaging Protocol uses Lamport timestamps
and timestamp watermarks to determine when all intended destinations
have received a message.  In particular, it uses the global variables
shown at the right of Figure~\ref{Variables}.  A message carries {\tt
  myGroupWatermark} in the {\tt back} field of its header, as a form
of acknowledgment, so that the remote group can safely discard, from
its buffers, messages sent to the group that carry timestamps less
than this group watermark.  The primary and the backups use the same
timestamp value for a message, which is essential for timestamp
watermark-based buffer management.

\renewcommand{\tabcolsep}{0.05in}
\renewcommand{\baselinestretch}{0.8}
\begin{figure}\hbox{
\begin{tabular}{ll}
&\parbox[t]{2.6in}{\footnotesize\bf \hspace*{0.6in}For Each Connection}\\[0.01in]
{\footnotesize\tt\bf msgSeqCount}&
\parbox[t]{2.6in}{\footnotesize A variable used to assign a
  message sequence number to each application message sent on the
  connection.}\\[0.01in]
{\footnotesize\tt\bf receivedUpToMsn}&
\parbox[t]{2.6in}{\footnotesize A variable used to store the 
  sequence number of the last message received on the connection
  without a gap.}\\[0.01in]
{\footnotesize\tt\bf sent list}&
\parbox[t]{2.6in}{\footnotesize A linked list that stores the application 
messages originated by a member.}\\[0.01in]
{\footnotesize\tt\bf received list}&
\parbox[t]{2.6in}{\footnotesize A linked list that stores incoming
  application messages received by a member.}\\[0.01in]
{\footnotesize\tt\bf nack list}&
\parbox[t]{2.6in}{\footnotesize A linked list that stores entries for
  the missing messages expected by this receiver and due to be
  negatively acknowledged.}\\ \\
\end{tabular}
\hspace*{0.0in}
\begin{tabular}{ll}
&\parbox[t]{2.15in}{\footnotesize\bf \hspace*{0.5in}For Buffer Management}\\[0.01in]
{\footnotesize\tt\bf myTimestamp}&
\parbox[t]{2.15in}{\footnotesize A Lamport clock used to timestamp the
  messages that are transmitted.}\\[0.01in]
{\footnotesize\tt\bf myTimestampWatermark}&
\parbox[t]{2.15in}{\footnotesize A timestamp such that this member has
  received all messages up to this timestamp for all connections in
  which it is involved.}\\[0.01in]
{\footnotesize\tt\bf myGroupWatermark}&
\parbox[t]{2.15in}{\footnotesize The minimum of the timestamp
   watermarks of the members of the group.}\\ \\ \\ \\ \\ \\
\end{tabular}}
\vspace*{-0.05in}
\caption{Variables used for each connection and global variables used for buffer management.}
\label{Variables}
\end{figure}
\renewcommand{\baselinestretch}{1.5}

 Figure \ref{data-structures} shows the pseudocode for the {\tt
   OrderInfo} struct (lines 1-7) and the {\tt MsgOrder} struct (lines
 8-15).  The opaque field (line 12) stores different {\tt OrderInfo}
 entries for different message types.  For a normal read, and a
 sucessful nonblocking write, it stores the offset with respect to the
 lower bound {\tt m\_msgSeqNum} so that the upper bound can be
 calculated, which allows the merger of {\tt OrderInfo} entries for
 consecutively sent/delivered messages from the same connection into a
 single {\tt OrderInfo} entry.  For a nonblocking write, it stores the
 number of times that the same message has been attempted, but failed,
 to send. If other {\tt OrderInfo} entries have been created in
 between, there might be multiple {\tt OrderInfo} entries for the same
 message.  There is no {\tt OrderInfo} entry for a blocking write.

\vspace*{-0.1in}
\subsection{Reliable Message Delivery}
\vspace*{-0.05in}
Reliable message delivery is described below in terms of a {\tt
  Request} from a client group $C$ to a server group $S$.  The same
considerations apply for a {\tt Reply} from a server group $S$ to a
client group $C$.  The pseudocode for the Messaging Protocol is given
in Figure \ref{messaging-alg}.  

The primary in group $C$ multicasts messages originated by the
application to a destination group over a virtual connection.  A
backup in group $C$ creates and logs (but does not multicast) messages
originated by the application.  Restricting the actions of the backup
in this way reduces the amount of network traffic.

When the primary in group $C$ first multicasts an application message
to group $S$ on a connection, it stores the message in a {\tt sent
  list} for the connection (lines 15-17). The primary retransmits a
message in the {\tt sent list} if it does not receive an
acknowledgment for the message sufficiently promptly (as determined
by a timeout) (lines 45-46).

The primary in group $S$ includes, in the header ({\tt ack} field) of
each application message it multicasts to group $C$ on a connection,
the message sequence number of the last application message it
received without a gap from the primary in group $C$ on that
connection (line 10).  If the primary in group $S$ does not have a
message to multicast on the connection sufficiently promptly (as
determined by a timeout), it multicasts a {\tt FirstAck} message
containing an {\tt ack} for the last application message it received
without a gap (lines 47-48).

The primary (and a backup) in group $S$ checks the {\tt precedence}
field of a message it receives to determine whether the message
originated in a competing membership whose primary has higher
precedence.  If so, it multicasts the message on the intra-group
connection to ensure that all other members have also received the
message, resets its state and rejoins the group (lines 18-20).

The primary (and a backup) in group $S$ adds the application messages
it receives on the connection to a {\tt received list} for the
connection (lines 25, 31), and updates the {\tt receivedUpToMsn}
variable (last message received without a gap) (line 32).  If the
replica detects a gap in the message sequence numbers (lines 22-25),
it creates placeholders for the missing messages, and adds
corresponding entries to a {\tt nack list}.  If the replica receives a
retransmitted message, and a placeholder for the message exists, it
replaces the placeholder with the message and, otherwise, discards the
retransmitted message (lines 26-30).

\begin{figure*}[t]
\vspace*{-0.5in}
\begin{center}
\footnotesize
\hbox{
\parbox{3.4in}{
\begin{tabbing}
\hspace*{0.1in}\=\hspace*{0.1in}\=\hspace*{0.1in}\=\hspace*{0.1in}\=\hspace*{0.1in}\=\kill
\>{\tt struct OrderInfo} \\[-0.1in]
\>\{ \\[-0.1in]
1 \>\>{\tt OrderType m\_orderType} \\[-0.1in]
2 \>\>{\tt SeqNumType m\_orderSeqNum} \\[-0.1in]
3 \>\>{\tt union} \\[-0.1in]
\>\>\{ \\[-0.1in]
4 \>\>\>{\tt MsgOrder m\_msgO} \\[-0.1in]
5 \>\>\>{\tt MutexOrder m\_mtxO} \\[-0.1in]
6 \>\>\>{\tt TimeOrder m\_timeO} \\[-0.1in]
7 \>\>\>{\tt SocketOrder m\_socO} \\[-0.1in]
\>\>\} \\[-0.1in]
\>\} 
\end{tabbing}
}
\hspace*{1.5in} 
\parbox{3.4in}{
\begin{tabbing}
\hspace*{0.1in}\=\hspace*{0.1in}\=\hspace*{0.1in}\=\hspace*{0.1in}\=\hspace*{0.1in}\=\kill
\>{\tt struct MsgOrder} \\[-0.1in]
\>\{ \\[-0.1in]
8 \>\>{\tt ViewNumType m\_primaryViewNum} \\[-0.1in] 
9 \>\>{\tt MsgType m\_msgType} \\[-0.1in]
10 \>\>{\tt ConnSeqNumType m\_connSeqNum} \\[-0.1in] 
11 \>\>{\tt short m\_sockFd} \\[-0.1in]
12 \>\>{\tt unsigned short m\_opaque} \\[-0.1in]
13 \>\>{\tt GrpIdType m\_remoteGrpId} \\[-0.1in]
14 \>\>{\tt SeqNumType m\_msgSeqNum} \\[-0.1in]
15 \>\>{\tt SeqNumType m\_orderSeqNum} \\[-0.1in]
\>\} \\[-0.1in] 
\end{tabbing}
}}
\vspace*{-0.15in}
\caption{Pseudocode for the {\tt OrderInfo} struct and the {\tt MsgOrder} struct.}
\label{data-structures}
\end{center}
\end{figure*}

A backup in group $C$ acknowledges a {\tt FirstAck} message it
receives with a {\tt SecondAck} message (lines 60-68). The backup
sends a {\tt SecondAck} message in response to receiving a {\tt
  FirstAck} message only if the backup application has generated the
message that the {\tt FirstAck} message acknowledges. If there is no
backup in group $C$, the primary in group $C$ carries out this
responsibility.  The primary in group $S$ stops retransmitting a {\tt
  FirstAck} message on receiving the corresponding {\tt SecondAck}
message (lines 70-71).

If a primary in group $C$ receives too many {\tt FirstAck} messages
from the primary in group $S$, acknowledging the last message the
primary in group $C$ sent, then the primary in group $S$ has not
received a {\tt SecondAck} from the backups in group $C$.
Consequently, the primary in group $C$ invokes the intra-group flow
control mechanisms to slow down, so that the backups in group $C$ can
catch up (lines 65-66).

If the primary in group $S$ does not have a message to multicast on an
inter-group connection sufficiently promptly (as determined by a
timeout) after it has stopped retransmitting the {\tt FirstAck}
message due to receiving the {\tt SecondAck} message, it multicasts a
{\tt KeepAlive} message on the connection to indicate the liveness of
the connection (lines 54-55).

If the primary (a backup) in group $S$ determines that it has not received
a message from the primary in group $C$ on a connection, it multicasts
a {\tt Nack} message on the remote (local) connection (lines 49-52). Such a
determination occurs if:
\begin{itemize}
\item The primary or a backup in group $S$ sees a gap in the message
  sequence numbers of the messages it received (line 24), or
\item A backup in group $S$ receives a {\tt SecondAck} message that
  contains an {\tt ack} for a message that the backup has not
  received (line 72), or
\item A backup in group $S$ receives a message from the primary in
  group $S$ that orders a message that the backup has not received.
\end{itemize}

The primary and each backup in group $S$ periodically exchange {\tt
  Heartbeat} messages on the intra-group connection (lines 56-59), so
that the primary knows that the backup has not failed (and vice versa)
and the buffer management mechanisms work properly.

\begin{figure*}[pt]
\vspace*{-0.5in}
\begin{center}
\footnotesize
\hbox{
\parbox{3.4in}{
\begin{tabbing}
\hspace*{0.1in}\=\hspace*{0.1in}\=\hspace*{0.1in}\=\hspace*{0.1in}\=\hspace*{0.1in}\=\kill
   \>{\bf On sending an application message {\tt M}}\\[-0.1in]
   \>begin\\[-0.1in]
1  \>\>{\tt msn} $\leftarrow$ msg seq number assigned to {\tt M} \\[-0.1in]
2  \>\>{\tt ack} $\leftarrow$ {\tt msn} of the last msg received without a gap \\[-0.1in]
3  \>\>{\tt wm} $\leftarrow$ group-wide timestamp watermark \\[-0.1in]
4  \>\>{\tt ts} $\leftarrow$ timestamp assigned to {\tt M} \\[-0.1in]
5  \>\>{\tt so} $\leftarrow$ source ordering information \\[-0.1in]
6  \>\>{\tt ro} $\leftarrow$ remote ordering information \\[-0.1in] 
7  \>\>if {\tt amPrimary()} then \\[-0.1in]
8  \>\>\>record send message order \\[-0.1in] 
9  \>\>{\tt M.setMsgSeqNum(msn)} \\[-0.1in]
10 \>\>{\tt M.setAckField(ack)} \\[-0.1in]
11 \>\>{\tt M.setBackField(wm)} \\[-0.1in]
12 \>\>{\tt M.setTimestamp(ts)} \\[-0.1in]
13 \>\>{\tt M.setPrecedence(precedence)} \\[-0.1in] 
   \>\> // precedence of primary \\[-0.1in] 
14 \>\>{\tt M.piggybackOrderInfo(so,ro)} \\[-0.1in]
15 \>\>if {\tt amPrimary()} then \\[-0.1in]
16 \>\>\>multicast {\tt M} to destination group \\[-0.1in]
17 \>\>append {\tt M} to sent list for retransmission \\[-0.1in]
   \>end \\[-0.1in] \\[-0.1in] 
   \>{\bf On receiving an application message {\tt M}} \\[-0.1in]
   \>begin \\[-0.1in]
18 \>\>if {\tt M.precedence} $>$ precedence of primary then \\[-0.1in]
   \>\>\{ \\[-0.1in]
19 \>\>\>multicast {\tt M} on the intra-group connection \\[-0.1in]
20 \>\>\>reset state and rejoin the group \\[-0.1in]
   \>\>\} \\[-0.1in]
   \>\>else \\[-0.1in]
   \>\>\{ \\[-0.1in]
21 \>\>\>{\tt msn} $\leftarrow$ next expected msg seq number \\[-0.1in]
22 \>\>\>if {\tt msn} $<$ {\tt M.getMsgSeqNum()} then \\[-0.1in]
   \>\>\>\{ \\[-0.1in]
23 \>\>\>\>create placeholders for missing messages \\[-0.1in]
24 \>\>\>\>append a {\tt Nack} to nack list \\[-0.1in]
25 \>\>\>\>append {\tt M} to received list \\[-0.1in]
   \>\>\>\} \\[-0.1in]
26 \>\>\>else if {\tt msn} $>$ {\tt M.getMsgSeqNum()} then \\[-0.1in]
   \>\>\>\{ \\[-0.1in]
27 \>\>\>\>if {\tt M} was missing then \\[-0.1in]
   \>\>\>\>\{ \\[-0.1in]
28 \>\>\>\>\>replace the placeholder with {\tt M} \\[-0.1in]
29 \>\>\>\>\>remove the {\tt Nack} from nack list \\[-0.1in]
   \>\>\>\>\} \\[-0.1in]
   \>\>\>\>else \\[-0.1in]
30 \>\>\>\>\>discard retransmitted {\tt M} \\[-0.1in]
   \>\>\>\} \\[-0.1in]
   \>\>\>else \\[-0.1in]
31 \>\>\>\>append {\tt M} to received list \\[-0.1in]
32 \>\>\>update {\tt receivedUpToMsn} \\[-0.1in]
33 \>\>\>handle piggybacked ordering information \\[-0.1in]
   \>\>\} \\[-0.1in]
   \> end \\[-0.1in] \\[-0.1in]
   \>{\bf On delivering an application message {\tt M}} \\[-0.1in]
   \>begin \\[-0.1in]
34 \>\>{\tt M} $\leftarrow$ first message in received list \\[-0.1in]
35 \>\>if not a placeholder for {\tt M} then \\[-0.1in]
   \>\>\{ \\[-0.1in]
36 \>\>\>if {\tt amPrimary()} then \\[-0.1in]
37 \>\>\>\> deliver {\tt M} and create a received message order \\[-0.1in]
38 \>\>\>if {\tt amBackup()} then \\[-0.1in]
   \>\>\>\{ \\[-0.1in]
39 \>\>\>\> find first msg order \\[-0.1in]
40 \>\>\>\>if found and msg order orders {\tt M} then \\[-0.1in]
41 \>\>\>\>\>deliver {\tt M} \\[-0.1in]
   \>\>\>\} \\[-0.1in]
42 \>\>\>if {\tt M} is delivered then \\[-0.1in]
43 \>\>\>\>move {\tt M} from received list to delivered list \\[-0.1in]
   \>\>\} \\[-0.1in]
   \>end \\[-0.1in] \\[-0.1in] \\[-0.1in]  
\end{tabbing}
}
\hspace*{0.7in} 
\parbox{3.4in}{ 
\begin{tabbing}
\hspace*{0.1in}\=\hspace*{0.1in}\=\hspace*{0.1in}\=\hspace*{0.1in}\=\hspace*{0.1in}\=\kill
   \>{\bf On periodic processing} \\[-0.1in]
   \>begin \\[-0.1in]
44 \>\>{\tt M\_sent} $\leftarrow$ message in the sent list \\[-0.1in]
45 \>\>if {\tt amPrimary()} and {\tt M\_sent} not acked then \\[-0.1in]
46 \>\>\>retransmit {\tt M\_sent} \\[-0.1in] 
47 \>\>if {\tt amPrimary()} and {\tt SecondAck} not received then \\[-0.1in]
48 \>\>\>retransmit a {\tt FirstAck} for last msg received without a gap \\[-0.1in]
49 \>\>if {\tt amPrimary()} and nack list not empty then \\[-0.1in]
50 \>\>\>retransmit {\tt Nack} to the remote group \\[-0.1in]
51 \>\>if {\tt amBackup()} and nack list not empty then \\[-0.1in]
52 \>\>\>retransmit {\tt Nack} to the local group \\[-0.1in]
53 \>\>deliver the first ordered message, if any, to the application \\[-0.1in]
   \>end \\[-0.1in] \\[-0.1in] 
   \>{\bf On expiration of the {\tt KeepAlive} timer for a connection} \\[-0.1in]
   \>begin \\[-0.1in]
54 \>\>\>multicast a {\tt KeepAlive} message on the connection \\[-0.1in]
55 \>\>\>reset the {\tt KeepAlive} timer for the connection \\[-0.1in]
   \>end \\[-0.1in] \\[-0.1in] 
   \>{\bf On expiration of the Heartbeat timer} \\[-0.1in]
   \>begin \\[-0.1in]
56 \>\>if {\tt amPrimary()} then \\[-0.1in]
57 \>\>\> multicast a {\tt Heartbeat} message to the backups \\[-0.1in] 
   \>\>else // backup \\[-0.1in]
58 \>\>\>transmit a {\tt Heartbeat} message to the primary \\[-0.1in] 
59 \>\>reset the {\tt Heartbeat} timer \\[-0.1in]
   \>end \\[-0.1in] \\[-0.1in] 
   \>{\bf On receiving a FirstAck message {\tt M\_FirstAck}} \\[-0.1in]
   \>begin \\[-0.1in]
60 \>\>{\tt M} $\leftarrow$ message in the sent list \\[-0.1in]  
61 \>\>{\tt num\_acked} $\leftarrow$ number of {\tt FirstAck} received for {\tt M} \\[-0.1in]
62 \>\>{\tt MAX\_ACK} $\leftarrow$ max number of {\tt FirstAck} for {\tt M} \\[-0.1in]  
63 \>\>find message {\tt M} corresponding to {\tt M\_FirstAck} \\[-0.1in]
64 \>\>if {\tt M} is found then \\[-0.1in]
   \>\> \{  \\[-0.1in]
65 \>\>\>if {\tt num\_acked} $>$ {\tt MAX\_ACK} then \\[-0.1in]
66 \>\>\>\>invoke intra-group flow control \\[-0.1in]
   \>\>\>else  \\[-0.1in]
67 \>\>\>\>if {\tt amBackup()} or {\tt amTheOnlyPrimary()} then \\[-0.1in]
68 \>\>\>\>\>multicast a {\tt SecondAck} for {\tt M} \\[-0.1in]
   \>\> \} \\[-0.1in]
   \>end \\[-0.1in] \\[-0.1in]
   \> {\bf On receiving a SecondAck message {\tt M\_SecondAck}} \\[-0.1in]
   \> begin \\[-0.1in]
69 \>\>find message {\tt M} corresponding to {\tt M\_SecondAck} \\[-0.1in]
70 \>\>if {\tt M} is found then \\[-0.1in]
71 \>\>\>{\tt M.stopRetransmitFirstAck()} \\[-0.1in]
   \>\> else \\[-0.1in]
72 \>\>\>append a {\tt Nack} to the nack list \\[-0.1in]
   \> end \\[-0.1in] \\[-0.1in] 
   \> {\bf On receiving a Nack message {\tt M\_Nack}} \\[-0.1in]
   \> begin \\[-0.1in]
73 \>\>find message {\tt M}, received or sent, corresponding to {\tt M\_Nack} \\[-0.1in]
74 \>\>if {\tt M} is found then \\[-0.1in]
75 \>\>\>retransmit {\tt M} \\[-0.1in]
   \> end \\[-0.1in] \\[-0.1in]
   \> {\bf On garbage collecting an application message {\tt M}} \\[-0.1in]
   \> begin \\[-0.1in]
76 \>\>{\tt ts} $\leftarrow$  {\tt M.getTimestamp()} \\[-0.1in]
77 \>\>{\tt myGrpWM} $\leftarrow$  group-wide watermark \\[-0.1in]
78 \>\>{\tt remoteGrpWM} $\leftarrow$ remote group watermark \\[-0.1in]
79 \>\>if {\tt M} is on sent list then \\[-0.1in]
80 \>\>\>if {\tt ts} $<=$ {\tt remoteGrpWM} then \\[-0.1in]
81 \>\>\>\>remove {\tt M} from sent list and delete {\tt M} \\[-0.1in]
82 \>\>if {\tt M} is on delivered list then \\[-0.1in]
83 \>\>\>if {\tt ts} $<=$ {\tt myGrpWM} then \\[-0.1in]
84 \>\>\>\>remove {\tt M} from received list and delete {\tt M} \\[-0.1in]
   \> end \\[-0.1in]
\end{tabbing}
}}
\vspace*{-0.15in}
\caption{Pseudocode for the Messaging Protocol.}
\label{messaging-alg}
\end{center}
\end{figure*}

\vspace*{-0.1in}
\subsection{Total Ordering within Groups}
\vspace*{-0.05in}
The primary in a group communicates ordering information to the
backups in its group, so that they obtain the same results in the same
order as the primary and, thus, maintain strong replica consistency.

In particular, the primary in group $C$ piggybacks, on each message it
originates and sends on a connection, the ordering information for the
messages it sent on the connection and received on the connection
since the last message it sent on the connection (along with ordering
information for other types of operations, as described in Section
\ref{determinizer-sec}).  A backup in group $C$ does not receive the
ordering information directly from the primary in group $C$.  Instead,
the primary in group $S$ reflects back the ordering information to
group $C$ in the next message it multicasts to group $C$.  The primary
in group $C$ includes the ordering information in each message it
sends until it receives that information reflected back to it.
Similarly, the primary in group $C$ reflects back to group $S$ the
ordering information it receives in messages from the primary in group
$S$.

\vspace*{-0.1in}
\subsection{Buffer Management}
\vspace*{-0.05in}
A replica must retain each message that it originates and that it
receives, until it knows that it will no longer need the message,
either to retransmit the message in response to a negative
acknowledgment or, to process the message if the primary fails and it
becomes the new primary.

In LLFT, timestamps and timestamp watermarks are used for buffer
management. Each replica in a group maintains a timestamp {\tt
  myGroupWatermark}.  Each message carries in the {\tt back} field of
the message header the group watermark of the sending
group. Each replica in a group maintains, for each connection, a {\tt
  remoteGroupWatermark}, to store the latest group watermark received
from the remote group of that connection.

As shown in Figure \ref{messaging-alg} (lines 76-84), a replica that
sends a message garbage collects the message if the {\tt timestamp} in
the message header is less than or equal to the {\tt
  remoteGroupWatermark} for the connection on which the message is
sent.  A replica that receives and delivers a message garbage collects
the message if the {\tt timestamp} in the message header is less than
or equal to the {\tt myGroupWatermark} for that replica's group.

\vspace*{-0.1in}
\section{Leader-Determined Membership Protocol}
\vspace*{-0.05in}
The formation of a membership has been based on two-phase commit with
a majority of correct processes to achieve consensus agreement and
avoid the split brain problem in which two or more competing
memberships are formed \cite{PaxosACMTrans,Paxos,Liskov}.
Unfortunately, in the presence of unreliable communication, it is
difficult or expensive to eliminate the risk of competing memberships.
If a communication problem occurs, some of the members might form a
new membership, while other members continue to operate with the
existing membership.  It is possible to avoid such a situation only if
every value that is communicated is subjected to a majority vote of
the membership, which is what is done in aircraft flight control
systems \cite{SIFT}.  Under conditions of unreliable communication, it
is undesirable to degenerate into multiple competing memberships,
possibly singleton memberships, and it is also undesirable to fail to
form a membership.  The objective must be to form the best possible
membership (a heuristic criterion), to detect and heal partitions that
form, and to reestablish a consistent state following recovery from a
partition \cite{NetworkPartitioning}.

The LLFT Membership Protocol addresses the problem of maintaining a
consistent view of the membership at the primary and the backups. It
ensures that they have the same membership set, the same primary view
number, and the same primary, by handling changes at the primary and
the backups.  The primary, in turn, determines the addition (removal)
of the backups to (from) the group, as well as their precedences and
ranks (defined below).  By making a deterministic choice of the
primary, the Membership Protocol is faster than a multi-round
consensus algorithm \cite{Chandra:UnrelFD}, which is particularly
important when normal processing is suspended by primary failure.

The {\it precedence} of a member of a group is determined by the order
in which the member joins the group.  If a member fails and
subsequently restarts, it is considered a new member, and is assigned
a new precedence. The precedences increase monotonically so that, in
the infinite sequence of consecutive primary views for a group, no two
members have the same precedence and a member that joins later has a
higher precedence.  When a primary adds a new backup to the
membership, it assigns the next precedence in sequence to that backup.
The precedences of the members determine the order of succession to
become the new primary, if the primary fails.

The {\it rank} of the primary member of a group is $1$, and the ranks
of the backup members are $2, 3, \ldots$ When a proposed new primary
assigns ranks to the backups of a new membership or when a primary
adds a new backup to the membership, it assigns those ranks in the
order of their precedences.  The ranks determine the timeouts for
detection of faults in the primary or the backups.  

The rank of a member can change when another member is removed from
the group, whereas the precedence of a member is assigned when it
joins the group and does not change while it is a member.  The ranks
of the members are consecutive, whereas the precedences need not be,
due to removal of a member from the group.

To avoid the situation where two backups both claim to be the next new
primary, the fault detection timeouts for the backups increase with
increasing rank.  The backup with rank 3 operates a fault detector to
determine that the primary is faulty and that the backup with rank 2
is faulty, because it did not determine that the primary is faulty.
Thus, the fault detector operated by the backup with rank 3 has a
longer timeout than the fault detector operated by the backup with
rank 2, {\it etc.}

For efficiency reasons, the fault detector timeouts must be chosen
carefully. Timeouts that are too long cause unnecessary delays after a
fault, whereas timeouts that are too short cause membership churn and
readmission of members to the group.  For example, the timeout of the
fault detector operated by the backup with rank $2$ might be 10ms, the
timeout of the fault detector operated by the backup with rank $3$
might be 30ms, {\it etc}.  Thus, the fault detector timeout of the
backup with rank $3$ allows for 10ms of inaction by the primary, an
additional 10ms of inaction by the backup with rank $2$, and an
additional 10ms for skew between the timeouts of the backups.  Given
its longer timeout, timing out the backup with rank $3$ is rare.

However, it might still happen that two backups both propose to become
the new primary. In such a case, the backup with the lower precedence
gives up and the backup with the higher precedence continues.  For
example, if the backup with rank 2 and the backup with rank 3 both
propose to become the new primary, because the backup with rank 3 has
higher precedence, it overrides the backup with rank 2.

Only membership changes that correspond to a change of the primary
constitute a new view, which we refer to as a {\it primary view
  change}.  Each new primary view has a {\it primary view number}.
When the primary view changes, the proposed new primary adjusts the
members' ranks and resets the message sequence number to one on each
of its connections.

It is important for the backups to change the primary view at the same
virtual synchrony point as the primary. To this end, the new primary
produces an ordering information entry for the primary view change and
multicasts that entry to the backups, just like the other ordering
information.  A backup changes to the new primary view when it has
performed all of the operations that were ordered before the virtual
synchrony point, as described below.

Both the primary and the backups in a group need to know when there is
a change in the primary of the group, because at that point their
ranks change and the message sequence numbers are reset to one. They
also need to know about the addition (removal) of a backup, because
such an event can result in a change in their ranks.  Moreover, to
achieve reliable message delivery, both the primary and the backups
need to know when there is a change in the primary of a remote
group. However, they do not need to know about membership changes due
to addition (removal) of a backup to (from) the remote group.

\vspace*{-0.1in}
\subsection{Data Structures}
\vspace*{-0.05in} 
\subsubsection{{\bf Message Types}}
\begin{figure}[t]
\begin{center} 
\leavevmode
\epsfxsize=3.1in 
\vspace{-0.2in} 
\epsfbox{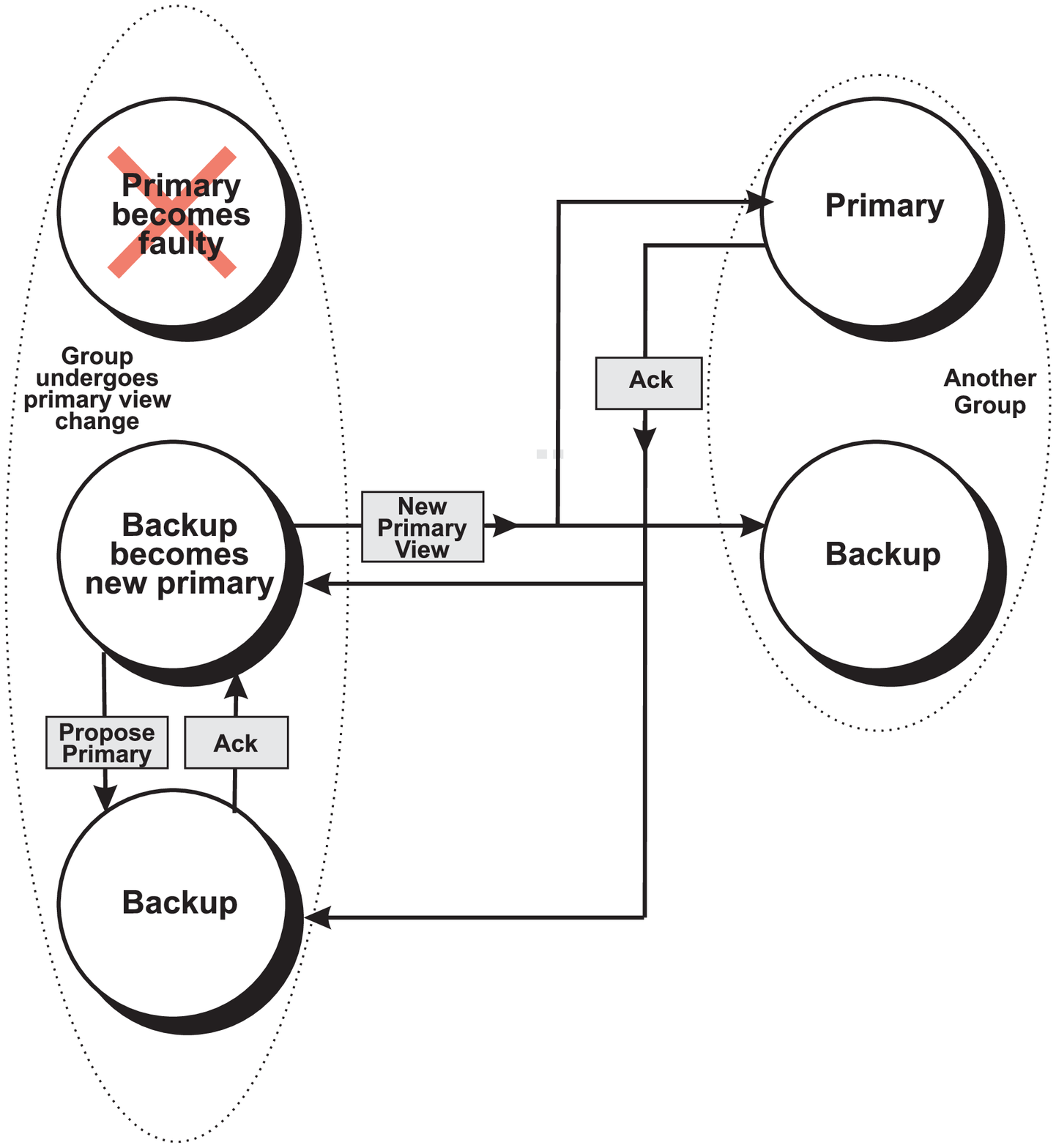}
\caption{Message exchange when a primary view change occurs.}
\vspace{-0.15in}
\label{reliable-membership}
\end{center}
\end{figure}

The types of messages used by the Membership Protocol are described in
Figure \ref{MembershipMessages} and are illustrated in Figure
\ref{reliable-membership}.

\renewcommand{\tabcolsep}{0.05in}
\renewcommand{\baselinestretch}{0.8}
\begin{figure}\hbox{
\begin{tabular}{ll}
&\parbox[t]{2.5in}{\footnotesize\bf \hspace*{0.15in}Message Types for Primary Change}\\[0.01in]
{\footnotesize\tt\bf ProposePrimary}&
\parbox[t]{2.5in}{\footnotesize A message multicast by a self-appointed
   new primary to request a change of the primary.}\\[0.01in]
{\footnotesize\tt\bf NewPrimaryView}&
\parbox[t]{2.5in}{\footnotesize A message multicast by the new
   primary on each of its connections (other than the intra-group
   connection) to report the new primary view and to collect
   information regarding the old primary.}\\[0.01in] \\[0.01in] \\[0.01in]
\end{tabular}
\hspace*{0.0in}
\begin{tabular}{ll}
&\parbox[t]{2.6in}{\footnotesize\bf \hspace*{0.3in}Message Types for Backup Change}\\[0.01in]
{\footnotesize\tt\bf ProposeBackup}&
\parbox[t]{2.6in}{\footnotesize A message multicast by a new replica
   that wants to join the group.}\\[0.01in]
{\footnotesize\tt\bf AcceptBackup}&
\parbox[t]{2.6in}{\footnotesize A message multicast by the primary to add
   a backup to the group.}\\[0.01in]
{\footnotesize\tt\bf RemoveBackup}&
\parbox[t]{2.6in}{\footnotesize A message multicast by the primary to
   remove a backup from the group.}\\[0.01in] 
{\footnotesize\tt\bf State}&
\parbox[t]{2.6in}{\footnotesize A message sent by the primary to a
  backup, containing the checkpointed state of the primary.} \\[0.01in]
\end{tabular}}
\caption{The types of messages used by the Membership Protocol for change of the primary and addition/removal of a backup.}
\label{MembershipMessages}
\end{figure}
\renewcommand{\baselinestretch}{1.5}

The {\tt ProposePrimary}, {\tt ProposeBackup}, {\tt AcceptBackup} and
{\tt RemoveBackup} messages are multicast on the intra-group
connection.

The {\tt ProposePrimary}, {\tt AcceptBackup} and {\tt RemoveBackup}
messages include the old membership in the payload, and require an
explicit acknowledgment from each backup.  For the primary, these
acknowledgment messages serve as ``commit'' messages.  The primary
(including the self-appointed new primary) must retransmit these
messages until all of the backups in the membership (as determined by
the primary) have acknowledged them.  The reason is that all of the
members in the group must have a consistent view of the membership and
the ranks of the members.

\vspace*{-0.1in}
\subsection{Change of the Primary}
\vspace*{-0.05in}
The change of the primary in a group is handled in two phases, as
described below.  The pseudocode for the Membership Protocol for
change of the primary is shown in Figure \ref{alg-primary-change}.  In
the rules below, $V_i$ denotes the primary view with primary view
number $i$ which corresponds to {\tt myPvn} in the pseudocode, and $p$
denotes the precedence of the primary which corresponds to {\tt
  myPrecedence} in the pseudocode.

\subsubsection{{\bf Determining the New Membership}}
In the first (election) phase, the new primary is determined. The new
primary determines which backups are included in the new membership
and their precedences and ranks.  More specifically, the first phase
operates as follows:
\begin{itemize}
\item If a backup with precedence $p$ does not receive a {\tt
  Heartbeat} message from the primary of view $V_i$ within a given
  time period (and thus determines that the primary is faulty) and it
  has not received a {\tt ProposePrimary} message for view $V_i$ from
  a backup with precedence $< p$, the backup multicasts a {\tt
    ProposePrimary} message on the intra-group connection, denouncing
  the old primary and appointing itself as the new primary of view
  $V_{i+1}$.
\begin{itemize}
 \item The backup excludes from the membership the old primary and the backups
   with precedences $< p$ (line 4). It excludes such a backup because that backup
   did not send a {\tt ProposePrimary} message quickly enough to
   become the new primary and, thus, is declared to be faulty.
\item The backup includes, in the {\tt ProposePrimary} message, the
  group identifier, the proposed new membership, its current primary
  view number $i$ and its precedence $p$ (line 5).
\end{itemize}
 \item If a backup with precedence $q$ receives a {\tt ProposePrimary}
   message for a new primary view $V_{i+1}$, from a proposed new
   primary with precedence $p$, and the backup is included in the
   proposed new membership (which implies that $q > p$), and 
\begin{itemize}
\item The backup has not generated a {\tt ProposePrimary} message for 
  view $V_{i+1}$, and
\item The backup has not acknowledged a {\tt ProposePrimary} message
  from a backup with precedence $> p$ for view $V_{i+1}$
\end{itemize}
   then the backup with precedence $q$ accepts the proposed new
   membership and acknowledges the {\tt ProposePrimary} message (lines
   21-24).
\item If a backup receives a {\tt ProposePrimary} message for new
  primary view $V_{i+1}$, or a subsequent view, from a proposed new
  primary with precedence $p$, and the backup is not included in the
  proposed new membership, and 
\begin{itemize}
\item The backup has not generated a {\tt
  ProposePrimary} message for view $V_{i+1}$ and $q > p$, and
\item The backup with precedence $q$ has not received a {\tt
  ProposePrimary} message for view $V_{i+1}$ from a backup with precedence $> p$
\end{itemize}
then the backup resets its state and rejoins the group (line 25).
 \item When the proposed new primary has received acknowledgments for
   its {\tt ProposePrimary} message from all members in the proposed
   new membership, it concludes the first phase and proceeds to the
   second phase (lines 14-16).
\end{itemize}
Note that the sets of conditions in the second and third bullets above
are not complementary and collectively exhaustive.  If a backup
receives a {\tt ProposePrimary} message that does not satisfy either
of those sets of conditions, it ignores that {\tt
  ProposePrimary} message.  The mechanisms for change of the primary determine
the new membership of the group using only one round of message
exchange ({\tt ProposePrimary} and corresponding acknowledgments).  In
a tradeoff for simplicity and timeliness, the mechanisms do not attempt
to form a new membership with the maximum possible number of members.

\begin{figure*}[t]
\vspace*{-0.5in}
\begin{center}
\footnotesize
\hbox{
\parbox{3.4in}{
\begin{tabbing}
\hspace*{0.1in}\=\hspace*{0.1in}\=\hspace*{0.1in}\=\hspace*{0.1in}\=\hspace*{0.1in}\=\hspace*{0.1in}\=\hspace*{0.1in}\=\kill
\>{\bf On expiration of the fault detection timeout for the primary} \\[-0.1in]
\>(at a backup) \\[-0.1in]
\>begin \\[-0.1in]
 1 \>\>{\tt myPvn} $\leftarrow$ current primary view number \\[-0.1in]  
 2 \>\>{\tt myPrecedence} $\leftarrow$ precedence assigned to this member \\[-0.1in]
 3 \>\>if not received a {\tt ProposePrimary} message {\tt M} such that \\[-0.1in]
   \>\>\>\>{\tt M.pvn} $>=$ {\tt myPvn} and {\tt M.precedence} $<$ {\tt myPrecedence} then \\[-0.1in] 
 \>\>\{ \\[-0.1in]
 4 \>\>\>exclude all members with lower precedences than {\tt myPrecedence} \\[-0.1in]
 5 \>\>\>multicast a {\tt ProposePrimary} message {\tt Mp} with the group id, \\[-0.1in]
   \>\>\>\>\>the new membership, {\tt myPvn} and {\tt myPrecedence} \\[-0.1in]
 6 \>\>\>start retransmission timer \\[-0.1in]
 7 \>\>\>{\tt retransmissionCount} $\leftarrow$ 0 \\[-0.1in]
   \>\>\} \\[-0.1in]
\>end \\[-0.1in] \\[-0.1in]
\>{\bf On expiration of the retransmission timer} (at the backup \\[-0.1in]
\>that sent the {\tt ProposePrimary} message {\tt Mp}) \\[-0.1in]
\>begin \\[-0.1in]
8 \>\>if {\tt retransmissionCount} $>$ {\tt MAX\_COUNT} then \\[-0.1in]
   \>\>\{ \\[-0.1in]
9 \>\>\>exclude members that have not yet acknowledged my \\[-0.1in]
  \>\>\>\>\>{\tt ProposePrimary} message {\tt Mp} \\[-0.1in]
10 \>\>\>transmit a {\tt ProposePrimary} {\tt Mp} with the latest membership \\[-0.1in]
11 \>\>\>{\tt retransmissionCount} $\leftarrow$ 0 \\[-0.1in]
   \>\>\} \\[-0.1in]
12 \>\>restart retransmission timer \\[-0.1in]
13 \>\>{\tt retransmissionCount}++ \\[-0.1in]
\>end \\[-0.1in] \\[-0.1in]
\>{\bf On receiving an ack for the {\tt ProposePrimary} message {\tt Mp}} \\[-0.1in] 
\>(at the backup that sent {\tt Mp}) \\[-0.1in]
\>begin \\[-0.1in]
14 \>\>if received acks from all backups in membership then \\[-0.1in]
\>\>\{ \\[-0.1in]
15 \>\>\>cancel retransmission timer for {\tt ProposePrimary} message {\tt Mp} \\[-0.1in]
16 \>\>\>start recovery protocol \\[-0.1in]
\>\>\} \\[-0.1in]
\>end \\[-0.1in] \\[-0.1in]
\>{\bf On receiving a {\tt ProposePrimary} message {\tt Mp}} (at a backup \\[-0.1in]
\>that did not send {\tt Mp} \\[-0.1in]
\>// {\tt primaryPrecedence} initially set to precedence of \\[-0.1in]
\>// primary in current view \\[-0.1in]
\>begin \\[-0.1in] 
17 \>\>{\tt myPvn} $\leftarrow$ current primary view number \\[-0.1in]\\[-0.1in]
\end{tabbing}
}
\hspace*{0.0in} 
\parbox{3.4in}{
\begin{tabbing}
\hspace*{0.1in}\=\hspace*{0.1in}\=\hspace*{0.1in}\=\hspace*{0.1in}\=\hspace*{0.1in}\=\hspace*{0.1in}\=\kill
18 \>\>if {\tt Mp.pvn} $>=$ {\tt myPvn} then \\[-0.1in]
19 \>\>\>if I am in the membership then \\[-0.1in]
   \>\>\>\{ \\[-0.1in]
20 \>\>\>\>if {\tt Mp.precedence} $>$ {\tt primaryPrecedence} then \\[-0.1in]
  \>\>\>\>\{ \\[-0.1in]
21\>\>\>\>\>{\tt primaryPrecedence} $\leftarrow$ {\tt Mp.precedence} \\[-0.1in]
22\>\>\>\>\>send acknowledgment for {\tt ProposePrimary} message \\[-0.1in]
23\>\>\>\>\>update membership and ranks \\[-0.1in]
24\>\>\>\>\>start fault detection timer for the new primary \\[-0.1in]
  \>\>\>\>\} \\[-0.1in]
   \>\>\>\} \\[-0.1in]
   \>\>\>else \\[-0.1in]
25 \>\>\>\>reset state and rejoin the group \\[-0.1in]
\>end \\[-0.1in] \\[-0.1in]                               
\>{\bf On recovering from a primary change} (at the new primary) \\[-0.1in]
\>begin \\[-0.1in]
26 \>\>{\tt myPvn}$++$ \\[-0.1in]
27 \>\>{\tt primaryPrecedence} $\leftarrow$ {\tt myPrecedence} \\[-0.1in]
28 \>\>for each connection do \\[-0.1in]
29 \>\>\>multicast a {\tt NewPrimaryView} message \\[-0.1in]
\>end \\[-0.1in] \\[-0.1in]
\>{\bf On receiving an ack for the {\tt NewPrimaryView} message} \\[-0.1in]
\>{\tt Mv} (at the new primary) \\[-0.1in] 
\>begin \\[-0.1in]
30 \>\>nack all missing messages until received them \\[-0.1in]
31 \>\>retrieve order info held at remote groups from application msg \\[-0.1in]
   \>\>\>\>or {\tt KeepAlive} msg \\[-0.1in]
32 \>\>if received all missing messages and reproduced \\[-0.1in] 
   \>\>\>\>all messages sent by the old primary then \\[-0.1in]
   \>\>\{ \\[-0.1in]
33 \>\>\>reset {\tt msgSeqNum} to 0 on each connection \\[-0.1in]
34 \>\>\>adjust ranks of backups \\[-0.1in]
35 \>\>\>record an order for the primary view change \\[-0.1in]
   \>\>\} \\[-0.1in]
\>end \\[-0.1in] \\[-0.1in]
\>{\bf On receiving a {\tt NewPrimaryView} message {\tt Mv}} (at the primary \\[-0.1in]
\>of a remote group) \\[-0.1in]
\>begin \\[-0.1in]
36 \>\>{\tt recvUpToMsn} $\leftarrow$ seq num of last msg received without a gap \\[-0.1in]
37 \>\>{\tt lastSentMsn} $\leftarrow$ seq num of last msg sent \\[-0.1in]
38 \>\>discard all messages received after {\tt recvUpToMsn} \\[-0.1in]
39 \>\>{\tt expectedPvn} $\leftarrow$ {\tt Mv.pvn} \\[-0.1in]
40 \>\>{\tt expectedMsn} $\leftarrow$ 0 \\[-0.1in]
41 \>\>acknowledge {\tt Mv} with ({\tt recvUpToMsn}, {\tt lastSentMsn}) \\[-0.1in]
\>end \\[-0.1in] 
\end{tabbing}
}}
\vspace*{-0.2in}
\caption{Pseudocode for the Membership Protocol to
handle the change of the primary.}
\label{alg-primary-change}
\end{center}
\end{figure*}
              
\subsubsection{{\bf Recovering from the Membership Change}}
In the second phase, the new primary queries the remote group of each
of its inter-group connections regarding the old primary's state, and
determines a virtual synchrony point.  The new primary needs to know
the last message sent by the old primary and delivered to each remote
group on a connection and, in particular, the ordering information
piggybacked onto the last message.  To advance to the state of the old
primary known to the remote groups before the old primary failed, the
new primary must follow the ordering information.  More specifically,
\begin{itemize}
\item The new primary collects information for the virtual synchrony
  point by multicasting a {\tt NewPrimaryView} message on each of its
  inter-group connections (lines 28-29).  The {\tt NewPrimaryView}
  message contains the most recent ordering information known to the
  new primary for the connection.
\item On receiving the {\tt NewPrimaryView} message, the primary of
  the remote group flushes all messages that came after the last
  message delivered from the old primary's group (line 38).  The
  primary of the remote group acknowledges the {\tt NewPrimaryView}
  message by providing information regarding the last message
  delivered from, and the last message sent to, the old primary's
  group (line 41). The primary of the remote group sends back the
  ordering information to the new primary either in a new application
  message, or in a {\tt KeepAlive} message if it does not have an
  application message to send.
\item On receiving an acknowledgment from the primary of the remote
  group, the new primary determines whether it has missed
  any messages from that primary.  The new primary then sends {\tt Nacks}
  for all missing messages until it has received them (line 30).  The
  new primary retrieves the ordering information piggybacked on
  application messages or {\tt KeepAlive} messages from the primary of
  the remote group.
\item When the new primary has executed all of the operations
  according to the ordering information determined by the old primary,
  it concludes the second phase by resetting the message sequence
  numbers to one, adjusting the backups' ranks, and generating an
  ordering information entry declaring the start of a new primary view
  (lines 33-35). The backups switch to the new primary view when they
  receive and process the ordering information.
\end{itemize}

\vspace*{-0.1in}
\subsection{Change of a Backup}
\vspace*{-0.05in}
The change of a backup is either the addition of a backup to the
group, or the removal of a backup from the group.  The pseudocode for
the Membership Protocol for the addition (removal) of a backup is
shown in Figure \ref{alg-backup-change}.  The pseudocode for joining a
process group (lines 1-14) includes the case where a process is the
first member of the group and, thus, is the primary.

\begin{figure*}[t]
\vspace*{-0.5in}
\begin{center}
\footnotesize
\hbox{
\parbox{3.4in}{
\begin{tabbing}
\hspace*{0.1in}\=\hspace*{0.1in}\=\hspace*{0.1in}\=\hspace*{0.1in}\=\hspace*{0.1in}\=\hspace*{0.1in}\=\hspace*{0.1in}\=\hspace*{0.1in}\=\hspace*{0.1in}\=\kill
\>{\bf On joining a process group} (at a new backup) \\[-0.1in]
\>begin \\[-0.1in]
 1 \>\>start logging \\[-0.1in]
 2 \>\>{\tt hostId} $\leftarrow$ host id of the joining process \\[-0.1in]
 3 \>\>{\tt pid} $\leftarrow$ process id of the joining process \\[-0.1in]
 4 \>\>{\tt ts} $\leftarrow$ local start up time of the joining process \\[-0.1in]
 5 \>\>{\tt myBirthId} $\leftarrow$ (hostId, pid, ts) \\[-0.1in]
 6 \>\>{\tt Mp} $\leftarrow$ {\tt ProposeBackup} message with {\tt myBirthId} \\[-0.1in]
 7 \>\>multicast {\tt ProposeBackup} message {\tt Mp} \\[-0.1in]
 8 \>\>start retransmission timer \\[-0.1in]
 9 \>\>{\tt retransmissionCount} $\leftarrow$ 0 \\[-0.1in]
\> end \\[-0.1in] \\[-0.1in]

\>{\bf On expiration of the retransmission timer for the} \\[-0.1in] 
\>{\bf {\tt ProposeBackup} message {\tt Mp}} (at the new backup) \\[-0.1in]
\>begin \\[-0.1in]
10 \>\>if {\tt retransmissionCount} $>$ {\tt MAX\_COUNT} then \\[-0.1in]
11 \>\>\>become the first member of the group and thus the primary \\[-0.1in]
   \>\>else \\[-0.1in]
   \>\>\{ \\[-0.1in]
12 \>\>\>retransmit {\tt ProposeBackup} message {\tt Mp} \\[-0.1in]
13 \>\>\>{\tt retransmissionCount}$++$ \\[-0.1in]
14 \>\>\>restart retransmission timer \\[-0.1in]
   \>\>\} \\[-0.1in]
\> end \\[-0.1in] \\[-0.1in]

\>{\bf On receiving an {\tt AcceptBackup} message {\tt Ma}} (at the new backup) \\[-0.1in]
\>begin \\[-0.1in]
15 \>\>if {\tt Ma.birthId} == {\tt myBirthId} then \\[-0.1in]
   \>\>\{ \\[-0.1in]
16 \>\>\>accept membership, precedence, rank \\[-0.1in]
17 \>\>\>cancel retransmission timer for {\tt ProposeBackup} message {\tt Mp} \\[-0.1in]
18 \>\>\>acknowledge {\tt AcceptBackup} message {\tt Ma} indicating \\[-0.1in]
   \>\>\>\>\>the need for a state transfer \\[-0.1in]
19 \>\>\>wait for a {\tt State} message \\[-0.1in]
   \>\>\} \\[-0.1in]
\> end \\[-0.1in] \\[-0.1in]

\>{\bf On receiving a {\tt State} message} (at the new backup) \\[-0.1in]
\>begin \\[-0.1in]
20 \>\>restore state \\[-0.1in]
21 \>\>replay messages from the log \\[-0.1in]
\> end \\[-0.1in] \\[-0.1in]

\>{\bf On receiving a {\tt ProposeBackup} message {\tt Mp}} (at the primary) \\[-0.1in]
\>begin \\[-0.1in]
22 \>\>if {\tt ProposeBackup} message {\tt Mp} is not a duplicate then \\[-0.1in]
   \>\>\{ \\[-0.1in]
23 \>\>\>assign precedence and rank \\[-0.1in]
24 \>\>\>add the backup to the membership \\[-0.1in]
25 \>\>\>execute commit protocol code \\[-0.1in]
26 \>\>\>start fault detection timer for the new backup \\[-0.1in]
   \>\>\} \\[-0.1in]
\>end 
\end{tabbing}
}

\hspace*{0.1in} 
\parbox{3.4in}{
\begin{tabbing}
\hspace*{0.1in}\=\hspace*{0.1in}\=\hspace*{0.1in}\=\hspace*{0.1in}\=\hspace*{0.1in}\=\hspace*{0.1in}\=\hspace*{0.1in}\=\kill
\>{\bf On expiration of the fault detection timer for a backup} \\[-0.1in]
\>(at the primary) \\[-0.1in]
\>begin \\[-0.1in]
27 \>\>remove the backup from the membership \\[-0.1in]
28 \>\>adjust ranks of the other backups \\[-0.1in]
29 \>\>execute commit protocol code \\[-0.1in]
\>end \\[-0.1in] \\[-0.1in]         

\>{\bf On committing a new membership} (at the primary) \\[-0.1in]
\>begin \\[-0.1in]
30 \>\>multicast a membership change message \\[-0.1in]
   \>\>// {\tt AcceptBackup} for adding a backup \\[-0.1in]
   \>\>// {\tt RemoveBackup} for removing a backup \\[-0.1in]
31 \>\>start retransmission timer \\[-0.1in]
32 \>\>{\tt retransmissionCount} $\leftarrow$ 0 \\[-0.1in]
\>end \\[-0.1in] \\[-0.1in]         

{\bf On expiration of the retransmission timer for the membership} \\[-0.1in]
\>{\bf change message} (at the primary) \\[-0.1in]
\>begin \\[-0.1in]
33 \>\>if {\tt retransmissionCount} $>$ {\tt MAX\_COUNT} then \\[-0.1in]
   \>\>\{ \\[-0.1in]
34 \>\>\>exclude members that have not yet acknowledged membership \\[-0.1in]
   \>\>\>\>\>change message \\[-0.1in]
35 \>\>\>retransmit membership change message \\[-0.1in]
   \>\>\>\>\>with latest membership \\[-0.1in]
36 \>\>\>{\tt retransmissionCount} $\leftarrow$ 0 \\[-0.1in]
   \>\>\} \\[-0.1in]
37 \>\>restart retransmission timer \\[-0.1in]
38 \>\>{\tt retransmissionCount}$++$ \\[-0.1in]
\>end \\[-0.1in] \\[-0.1in]         

{\bf On receiving an ack for the membership change message} \\[-0.1in]
\>(at the primary) \\[-0.1in]
\>begin \\[-0.1in]
39 \>\>if received acks for the membership change message from \\[-0.1in]
   \>\>\>\>all backups in membership then \\[-0.1in]
   \>\>\{ \\[-0.1in]
40 \>\>\>cancel retransmission timer for the membership change message \\[-0.1in]
41 \>\>\>get checkpoint of the state \\[-0.1in] 
42 \>\>\>send {\tt State} message to the backup \\[-0.1in]
   \>\>\} \\[-0.1in]
\>end \\[-0.1in] \\[-0.1in]

\>{\bf On receiving an {\tt AcceptBackup} / {\tt RemoveBackup}} \\[-0.1in] 
\>{\bf message} (at an existing backup) \\[-0.1in]
\>begin \\[-0.1in]
43 \>\>if I am in the membership then \\[-0.1in]
   \>\>\{ \\[-0.1in]
44 \>\>\>update the membership and ranks \\[-0.1in]  
45 \>\>\>send acknowledgment to primary \\[-0.1in]
   \>\>\} \\[-0.1in]
   \>\>else \\[-0.1in] 
46 \>\>\>reset state and rejoin the group \\[-0.1in]
\>end 
\end{tabbing}
}}
\vspace*{-0.15in}
\caption{Pseudocode for the Membership Protocol to
handle the addition and removal of a backup.}
\label{alg-backup-change}
\end{center}
\end{figure*}

\subsubsection{{\bf Addition of a Backup}}
A new process begins to log messages as soon as it starts up (line
1). The {\tt myBirthId} of a process (line 5) is a unique identifier,
similar to a birth certificate.  It is used to distinguish a process
that wishes to join the membership and that doesn't yet have a
precedence. The precedence is a unique identifier for a process only
after it is a member, and denotes the order in which it has become a
member.  The process multicasts a {\tt ProposeBackup} message on the
intra-group connection (line 7). The primary assigns the precedence
and the rank of the new backup (line 23) and then multicasts an {\tt
  AcceptBackup} message (line 30), containing the new membership, on
the intra-group connection. A backup that receives an {\tt
  AcceptBackup} message, that includes itself in the membership,
accepts the new membership, and responds with an acknowledgment (lines
15-18).

The primary checkpoints its state when it has received
acknowledgments for the new membership from all of the backups in the
group (lines 39-41). The point at which the checkpoint is taken
represents the virtual synchrony point for adding the new backup.  The
primary transmits the checkpoint to the new backup in a {\tt State}
message (line 42).  The new backup then sets its state by applying the
checkpoint, and replaying the messages from the log (lines 20-21),
after deleting obsolete messages.

\subsubsection{{\bf Removal of a Backup}}
The primary modifies the ranks of the backups in the group (line 28)
and then multicasts a {\tt RemoveBackup} message (line 30), containing
the new membership, on the intra-group connection.  When a backup
receives a {\tt RemoveBackup} message that includes itself in the
membership, the backup accepts the new membership and responds with an
acknowledgment (lines 43-45).  When a backup receives a {\tt
  RemoveBackup} message that does not include itself in the
membership, the backup resets its state and multicasts a {\tt
  ProposeBackup} message requesting to be readmitted to the membership
(line 46).

For both addition and removal of a backup, the primary multicasts the
new membership to all of the backups in the membership (line 30), and
asynchronously collects acknowledgments from all of them. It commits
the membership change when it has collected acknowledgments from all
of the backups in the membership (line 39). If a backup does not
provide an acknowledgment promptly, the primary removes the backup
from the membership (line 34).

\vspace*{-0.1in}
\section{Virtual Determinizer Framework}
\label{determinizer-sec}
\vspace*{-0.05in}
A reliable, totally ordered message delivery protocol ensures consistent
replication only if the application is deterministic (or is rendered
deterministic).  However, modern applications are typically
non-deterministic in a number of ways.  To maintain strong replica
consistency, it is necessary to sanitize or mask such sources of
non-determinism, {\it i.e.}, to render the application {\it virtually
  deterministic}.

The LLFT Virtual Determinizer Framework introduces a novel generic
algorithm for sanitizing the sources of non-determinism in an
application in a transparent manner.  We describe the generic
algorithm below, after describing the threading model.

\vspace*{-0.1in}
\subsection{Threading Model}
\vspace*{-0.05in}
The state of an application process is determined by data that are
shared among different threads, and by thread-specific local data managed
and changed by each thread.

Each thread within a process has a unique thread identifier.  A data
item that is shared by multiple threads is protected by a mutex.  The
threads and mutexes can be created and deleted dynamically.

Each replica in a process group runs the same set of threads.  A
thread interacts with other threads, processes, and its runtime
environment through system/library calls.  Non-determinism can arise
from different orderings of, and different results from, such calls at
different replicas in the group.

If the operations on the shared and local data in different replicas
are controlled in such a way that (1) the updates on a data item occur
in the same order with the same change, and (2) each thread updates
different data items in the same order with the same change, then the
replicas will remain consistent.

Figure \ref{determinizer} on the left gives example pseudocode for a
thread that shows how such calls might change the state of an
application. The pseudocode uses three types of system/library calls:
\begin{itemize}
\item Calls that try to acquire a mutex (line 18). The {\tt pthread\_}
  {\tt mutex\_trylock()} operation is similar to a nonblocking read in
  that, if the mutex is currently held by another thread, the call
  returns immediately with a specific error code, so that the caller
  thread is not blocked. If the thread of one replica successfully
  claims the mutex, while the corresponding thread of another replica
  fails, the two replicas perform different operations (lines 19-22),
  causing divergence of their states, because one replica changes the
  shared data {\tt SD1} (line 20) while the other replica changes the
  thread-local data {\tt LD5} (line 22).

\item Calls that retrieve local clock values (lines 1, 13). These
  calls change thread-local data ({\tt LD1})
  directly (lines 2, 14). If different replicas obtain different clock values, the
  replicas might arrive at different decisions (line 15) as to whether
  a timeout occurred. If one replica times out while the
  other does not, the states of the replicas will diverge because
  of the difference in thread-local data {\tt LD4} (line 16).
\item Calls that read (write) from (to) a socket asynchronously (lines
  3, 7, 12). If, for the same read operation, one replica
  successfully reads a message while the other does not, the states of
  the two replicas will differ in the thread-local data {\tt LD2} (line 5) and
  potentially {\tt LD3} (lines 9, 11). The consequence of different
  results for a nonblocking write call is similar.
\end{itemize}

\begin{figure*}[t]
\vspace*{-0.5in}
\begin{center}
\footnotesize
\hbox{
\parbox{3.4in}{
\begin{tabbing}
\hspace*{0.1in}\=\hspace*{0.1in}\=\hspace*{0.1in}\=\hspace*{0.1in}\=\hspace*{0.1in}\=\kill
 1 \>\>{\bf get current time} \\[-0.1in]
 2 \>\>// update thread-local data {\tt LD1} \\[-0.1in]
 3 \>\>{\bf do a nonblocking read from socket fd} \\[-0.1in]
 4 \>\>if picked up a message then \\[-0.1in]
   \>\>\{ \\[-0.1in]
 5 \>\>\>// update thread-local data {\tt LD2} \\[-0.1in]
 6 \>\>\>handle the message \\[-0.1in]
 7 \>\>\>{\bf do a nonblocking write to socket fd} \\[-0.1in]
 8 \>\>\>if failed to write the response then \\[-0.1in]
   \>\>\>\{ \\[-0.1in]
 9 \>\>\>\>// update thread-local data {\tt LD3} \\[-0.1in]
10 \>\>\>\>append to a queued message, if any \\[-0.1in]
   \>\>\>\} \\[-0.1in]
   \>\>\} \\[-0.1in]
   \>\>else \\[-0.1in]
   \>\>\{ \\[-0.1in]
11 \>\>\>// update thread-local data {\tt LD3} \\[-0.1in]
12 \>\>\>{\bf flush queued message, if any, to socket fd} \\[-0.1in]
   \>\>\} \\[-0.1in]
13 \>\>{\bf get current time} \\[-0.1in]
14 \>\>// update thread-local data {\tt LD1} \\[-0.1in]
15 \>\>if timed out then \\[-0.1in]
   \>\>\{ \\[-0.1in]
16 \>\>\>// update thread-local data {\tt LD4} \\[-0.1in]
17 \>\>\>call timeout handling routine \\[-0.1in]
   \>\>\} \\[-0.1in]
18 \>\>{\bf try to claim mutex {\tt Mtx}} \\[-0.1in]
19 \>\>if claimed mutex {\tt Mtx} then \\[-0.1in]
   \>\>\{ \\[-0.1in]
20 \>\>\>change shared data {\tt SD1} \\[-0.1in]
21 \>\>\>release mutex {\tt Mtx} \\[-0.1in]
   \>\>\} \\[-0.1in]
   \>\>else \\[-0.1in]
22 \>\>\>update thread-local data {\tt LD5} 
\end{tabbing}
}
\hspace*{0.75in} 
\parbox{3.4in}{
\begin{tabbing}
\hspace*{0.1in}\=\hspace*{0.1in}\=\hspace*{0.1in}\=\hspace*{0.1in}\=\hspace*{0.1in}\=\kill
\>{\bf On returning from a call} (at the primary) \\[-0.1in]
\>begin \\[-0.1in]
23 \>\>{\tt T} $\leftarrow$ thread identifier \\[-0.1in]
24 \>\>{\tt O} $\leftarrow$ operation identifier \\[-0.1in]
25 \>\>{\tt N} $\leftarrow$ operation count \\[-0.1in]
26 \>\>{\tt D} $\leftarrow$ operation metadata \\[-0.1in]
27 \>\>{\tt OrderInfo} $\leftarrow$ global queue to store order info \\[-0.1in]
28 \>\>append an entry {\tt (T, O, N, D)} to {\tt OrderInfo} \\[-0.1in]
\>end \\[-0.1in] \\[-0.1in] \\[-0.1in]

\>{\bf On receiving an order info entry {\tt (T, O, N, D)}} \\[-0.1in]
\>(at a backup) \\[-0.1in]
\>begin \\[-0.1in]
29 \>\>if {\tt O.OrderInfo} does not exist then \\[-0.1in]
30 \>\>\>create {\tt O.OrderInfo} \\[-0.1in]
31 \>\>append {\tt (T, N, D)} to {\tt O.OrderInfo} \\[-0.1in]
32 \>\>{\tt T1} $\leftarrow$ first entry in {\tt O.OrderedInfo} \\[-0.1in]
33 \>\>wake up {\tt T1} if it is blocked \\[-0.1in]
\>end \\[-0.1in] \\[-0.1in] \\[-0.1in]

\>{\bf On intercepting a call} (at a backup) \\[-0.1in]
\>begin \\[-0.1in]
34 \>\>{\tt T1} $\leftarrow$ identifier of the thread performing the call \\[-0.1in]
35 \>\>{\tt O1} $\leftarrow$ operation identifier of the call \\[-0.1in]
36 \>\>{\tt N1} $\leftarrow$ count for {\tt O1} for any thread \\[-0.1in]
37 \>\>get first entry {\tt (T, N, D)} of {\tt O1.OrderInfo} \\[-0.1in]
38 \>\>while {\tt (T, N, D)} not available or {\tt T1 != T} or {\tt N1 != N} do \\[-0.1in]
39 \>\>\>suspend {\tt T1} \\[-0.1in]
40 \>\>consume {\tt (T, N, D)} and remove it from {\tt O1.OrderInfo} \\[-0.1in]
41 \>\>return \\[-0.1in]
\>end \\[-0.1in] 
\end{tabbing}
}}
\vspace*{-0.15in} 
\caption{On the left, pseudocode for a thread.  The system/library calls that
  might change the state, or lead to a state change, are highlighted
  in bold.  On the right, pseudocode for the Virtual Determinizer to render
  the application virtually deterministic.}
\label{determinizer}
\vspace*{-0.0in} 
\end{center}
\end{figure*}

\vspace*{-0.1in}
\subsection{Generic Algorithm}
\vspace*{-0.05in}
The generic algorithm, shown in Figure \ref{determinizer} on the right, records
the ordering information and the return value information of non-deterministic
system/library calls at the primary, to ensure that the backups obtain
the same results in the same order.  For each non-deterministic
operation, the algorithm records the following information:
\begin{itemize}
\item {\tt\bf Thread identifier} - The identifier of the thread that is
  carrying out the operation.
\item {\tt\bf Operation identifier} - An identifier that represents one
  or more data items that might change during the operation or on
  completion of the operation.
\item {\tt\bf Operation count} - The number of operations
  carried out by a thread for the given operation identifier.
\item {\tt\bf Operation metadata} - The data returned from the
  system/library call. This metadata includes the {\tt out} parameters
  (if any), the return value of the call, and the error code (if
  necessary).
\end{itemize}

At the primary, the algorithm maintains a queue, the {\tt OrderInfo}
queue of four-tuples {\tt (T, O, N, D)}, where thread {\tt T} has
executed a call with operation identifier {\tt O} and with metadata
recorded in {\tt D}, and this is the {\tt N}th time in its execution
sequence that thread {\tt T} has executed such a non-deterministic
call.  The {\tt OrderInfo} queue spans different threads and different
operations.

At the primary, the algorithm appends an entry {\tt (T, O, N, D)} to
the {\tt OrderInfo} queue on return of the operation {\tt
  O} (lines 23-28).  The entries are transmitted to the backups
reliably and in order, using the piggybacking mechanism of the
Messaging Protocol.

At a backup, for each operation {\tt O}, the algorithm maintains an
{\tt O.OrderInfo} queue of three-tuples {\tt (T, N, D)}, in the order
in which the primary created them. When the backup receives the first
entry {\tt (T, O, N, D)} for operation {\tt O}, it creates the
{\tt O.OrderInfo} queue (lines 29-30).  After the entry is appended to
the queue, the algorithm awakens the first application thread in the
{\tt O.OrderInfo} queue if it is blocked (lines 31-33).

At a backup, when thread {\tt T} tries to execute operation {\tt
  O} as its {\tt N}th execution in the sequence, if {\tt (T, N, D)}
is not the first entry in the {\tt O.OrderInfo} queue, the algorithm
suspends the calling thread {\tt T} (lines 34-39).  It resumes a
thread {\tt T} that was suspended in the order in which {\tt (T, N,
  D)} occurs in the {\tt O.OrderInfo} queue, rather than the order in
which the thread was suspended or an order determined by the operating
system scheduler. It removes an entry {\tt (T, N, D)} from the {\tt
  O.OrderInfo} queue immediately before it returns control to the
calling thread {\tt T} after its {\tt N}th execution in the sequence
(lines 40-41).  The algorithm requires the ordering of all related
operations, {\it e.g.}, both claims and releases of mutexes.

We have instantiated the generic algorithm of the Virtual Determinizer
Framework for several major types of non-determinism, including
multi-threading, time-related operations and socket communication, as
discussed below.

\vspace*{-0.1in}
\subsection{Multi-Threading}
\vspace*{-0.05in}
The Consistent Multi-Threading Service (CMTS)
creates mutex ordering information at the primary, where the {\tt
  operation identifier} is the mutex $Mtx$. For the normal mutex claim call
({\tt pthread\_} {\tt mutex\_lock()} library call), the {\tt operation
  metadata} can be empty if the call is successful.  However, if the
normal mutex claim call returns an error code and for the nonblocking mutex
claim call ({\tt pthread\_mutex\_trylock()} library call), the {\tt
  operation metadata} is the return value.

At a backup, to process a mutex ordering information entry, the CMTS
examines the metadata. If the metadata contain an error code, the CMTS
returns control to the calling thread with an identical error status,
without performing the call. Otherwise, it delegates the mutex claim
operation to the original library call provided by the operating
system.  If the mutex is not currently held by another thread, the
calling thread acquires the mutex immediately.  Otherwise, the calling
thread is suspended and subsequently resumed by the operating system
when the thread that owns the mutex releases it.

The CMTS allows concurrency of threads that do not simultaneously
acquire the same mutex. Thus, it achieves the maximum possible degree
of concurrency, while maintaining strong replica consistency.

\vspace*{-0.1in}
\subsection{Time-Related Operations}
\vspace*{-0.05in}
The Consistent Time Service (CTS) ensures that clock readings at
different replicas are consistent.  For time-related system calls,
such as {\tt gettimeofday()} and {\tt time()}, the CTS creates time
ordering information at the primary, where the {\tt operation
  identifier} is the time source and the {\tt operation metadata} is
the clock value, or an error code if the call fails.

In addition to consistency for each clock reading, the CTS ensures
monotonicity of the clock as seen by the replicas in a group, even if
the primary fails \cite{cts}. With the CTS, the replicas see a {\it
  virtual group clock} that resembles the real-time clock. Each
replica maintains an offset to record the difference between its local
physical clock and the virtual group clock. The offset of the primary
is 0. Each backup updates its offset for each clock reading.

If the primary fails, one of the backups becomes the new primary. The
new primary must not include its local physical clock value in the
time ordering information it sends to the backups, because doing so
might roll backward, or roll forward, the virtual group
clock. Instead, the new primary adds the recorded offset to its local
physical clock value, and includes that value in the time ordering
information it sends to the backups.

\vspace*{-0.1in}
\subsection{Socket Communication}
\vspace*{-0.05in}
An application might use a nonblocking read to receive messages from
the network asynchronously.  If no message is received, the
nonblocking read call returns a specific error code.  On such an error
return, the application might switch to some other task and change to
a different state.  Thus, the event of failing to receive a message
must be ordered.  Similarly, an application might use a nonblocking
write to send data asynchronously. If the message is not sent
successfully, the application might take on a different task and
change to a different state. Thus, the event of failing to send a
message must be ordered.

On return from a read/write system call on a socket at the primary,
the Consistent Socket Communication Service (CSCS) produces a socket
ordering information entry for that operation. The {\tt operation
  identifier} is the socket file descriptor. The {\tt operation
  metadata} is an identifier for the message being read/written, if
the read/write succeeds, or an error code, if it fails.

It is quite common to combine socket read/write system calls with
select/poll system calls. Typically, the application performs a
read/write system call only if the select/poll system call indicates
that the corresponding socket is readable/writable.  The select/poll
system call offers a timeout parameter for the user to specify how
long the operating system can take to return from the call.

The CSCS produces a socket ordering information entry on returning
from a select/poll system call. The {\tt operation identifier} is the
socket file descriptor.  The {\tt operation metadata} contains the
number of events, the read/write/error mask, and the amount of time
left before the timeout (used on Linux) if the call returns
successfully, or an error code, if it fails.

\vspace*{-0.1in}
\section{Implementation and Performance}
\vspace*{-0.05in}
The LLFT system has been implemented in the C++ programming language
for the Linux operating system. The library interpositioning technique
is used to capture and control the application's interactions with its
runtime environment. Application state is checkpointed and restored
using facilities provided by a memory-mapped checkpoint library
derived from \cite{dieter::chkpt}. The implementation of LLFT is
compiled into a shared library. The library is inserted into the
application address space at startup time using the {\tt LD\_PRELOAD}
facility provided by the operating system. LLFT is transparent to the
application being replicated, and does not require recompilation or
relinking of the application program.

The experimental testbed consists of 14 HP blade servers, each
equipped with two 2GHz Intel Xeon processors, running the Ubuntu 9.04
operating system, on a 1Gbps Ethernet. A two-tier client/server
application was used to benchmark the LLFT implementation.  The
performance evaluation focuses on three areas: (1) performance of the
Messaging Protocol during normal fault-free operation, (2) overhead of
the Virtual Determinizer Framework, and (3) performance of the
Membership Protocol during fault recovery.

\begin{figure*}[t]
\begin{center}
\leavevmode
\hbox{\parbox{3.65in}{ 
\epsfxsize=3.6in 
\epsfbox{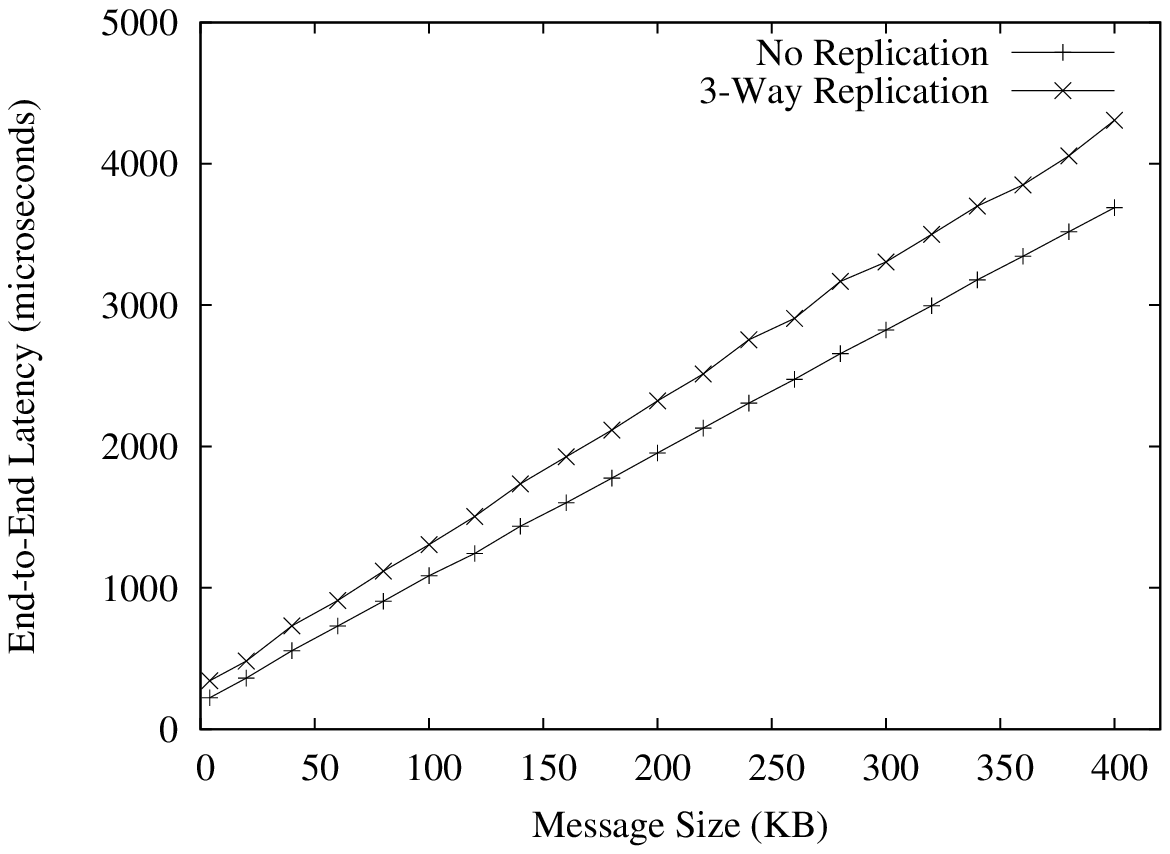}
\vspace{-0.15in} 
\caption{End-to-end latency vs. message size. }
\label{latency}
}
\hspace{0.05in} 
\parbox{3.65in}{
\epsfxsize=3.6in 
\epsfbox{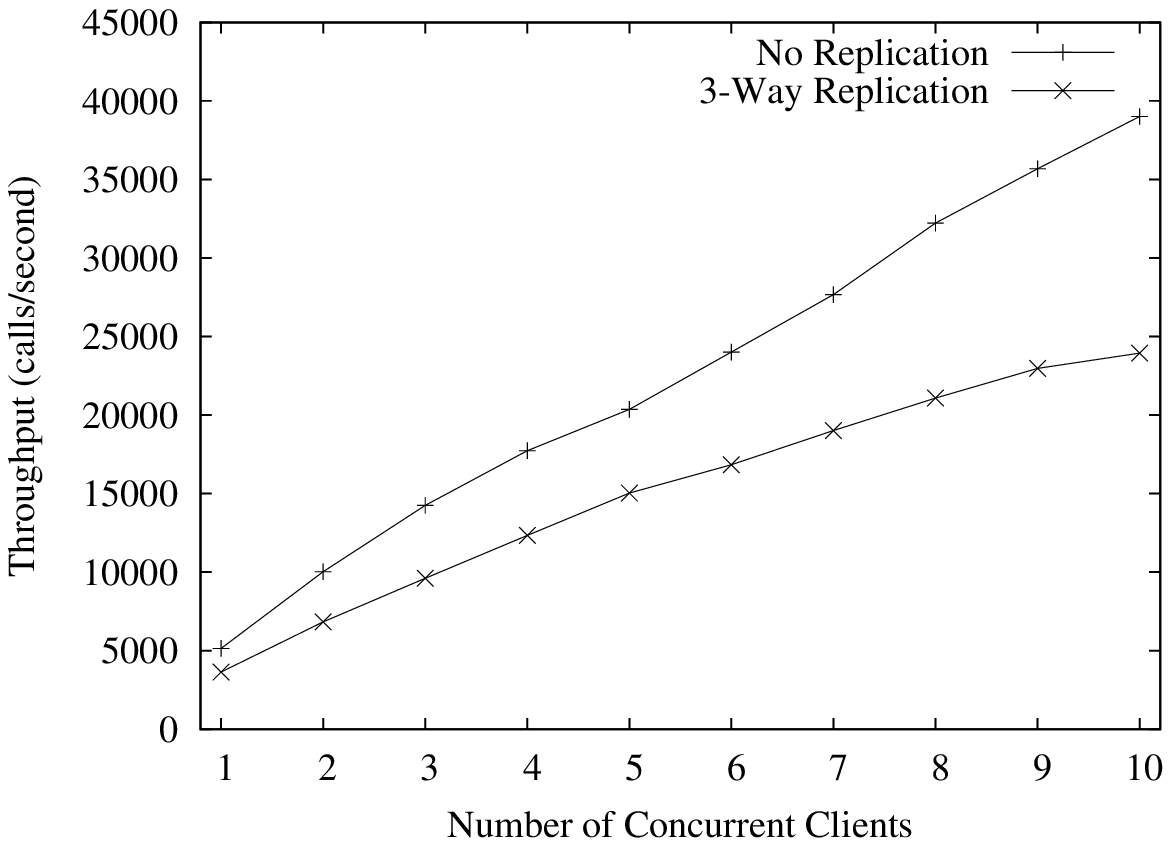}
\vspace{-0.15in}
\caption{Throughput vs. number of concurrent clients.}
\label{thrput}
}}
\vspace{-0.25in}
\end{center}
\end{figure*}

\vspace*{-0.1in}
\subsection{Messaging Protocol}
\vspace*{-0.05in}
First, we consider the performance of the Messaging Protocol during
normal fault-free operation. We characterize the end-to-end latency in
the presence of a single client for various invocation patterns: (1)
short requests and short replies, (2) various size requests and short
replies, and (3) short requests and various size replies. The
end-to-end latency for pattern (2) is virtually indistinguishable
from that for pattern (3) for the same message size (for requests and
replies). Hence, the measurement results shown in Figure~\ref{latency}
refer only to message size. The figure shows the end-to-end latency
without replication using TCP as a baseline for comparison and with
3-way replication using LLFT to understand the overhead that LLFT
incurs.  As can be seen in the figure, the Messaging Protocol incurs
very moderate overhead, ranging from about 15\% for large messages to
about 55\% for small messages. The overhead of the Messaging Protocol
is determined primarily by the piggybacking of ordering
information. For large messages, which require fragmentation in user
space, the Messaging Protocol incurs additional context switches,
although the relative overhead is less percentage-wise.

We also measured the throughput, without replication using TCP and
with 3-way replication using LLFT, in the presence of various numbers
of concurrent clients. Each client continually issues 1KB requests
without any think time, and the server responds with 1KB replies. The
measurement results are summarized in Figure~\ref{thrput}.  It can be
seen that, although the throughput reduction under replication is
moderate under light loads, it is more prominent under heavy
loads.

\begin{figure*}[t]
\begin{center}
\leavevmode
\hbox{\parbox{3.65in}{ 
\epsfxsize=3.6in 
\epsfbox{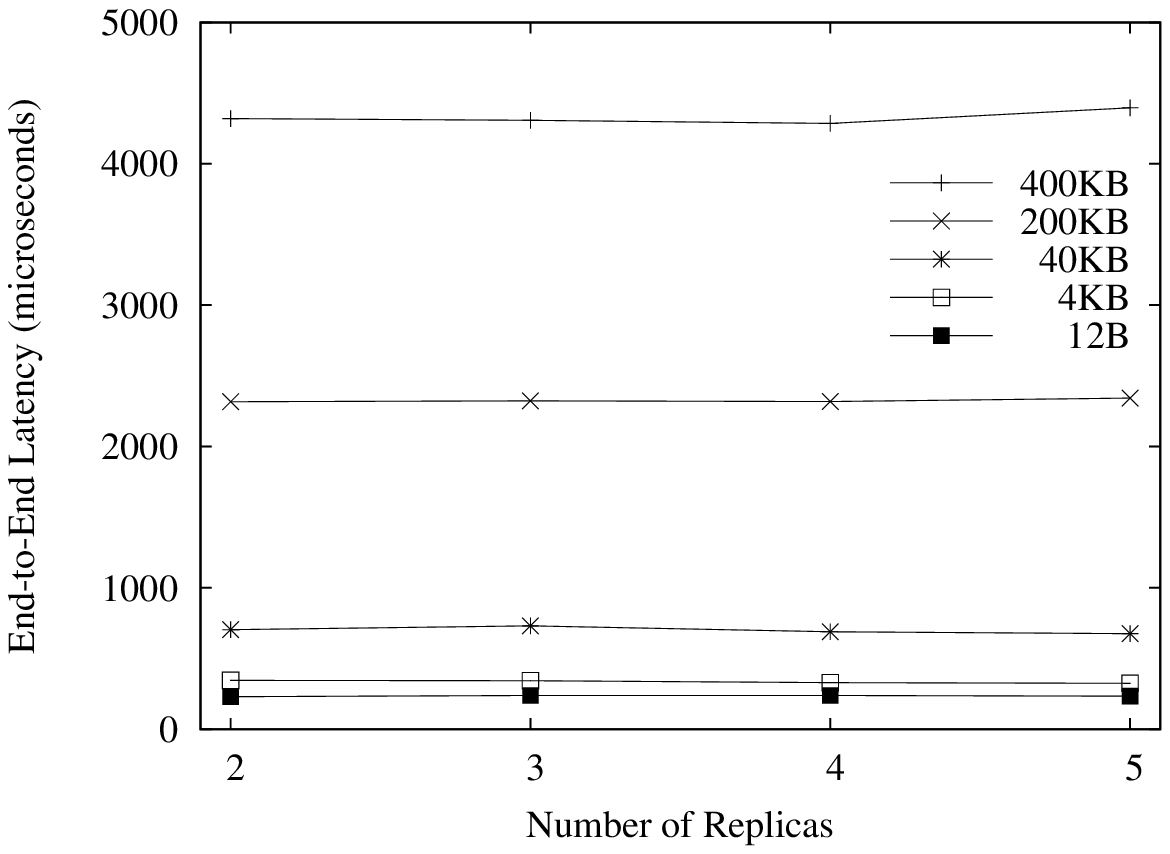}
\vspace{-0.15in}
\caption{End-to-end latency vs. number of replicas in a group.}
\label{scalabilityperf}
}
\hspace{0.15in}
\parbox{3.65in}{ 
\epsfxsize=3.6in 
\epsfbox{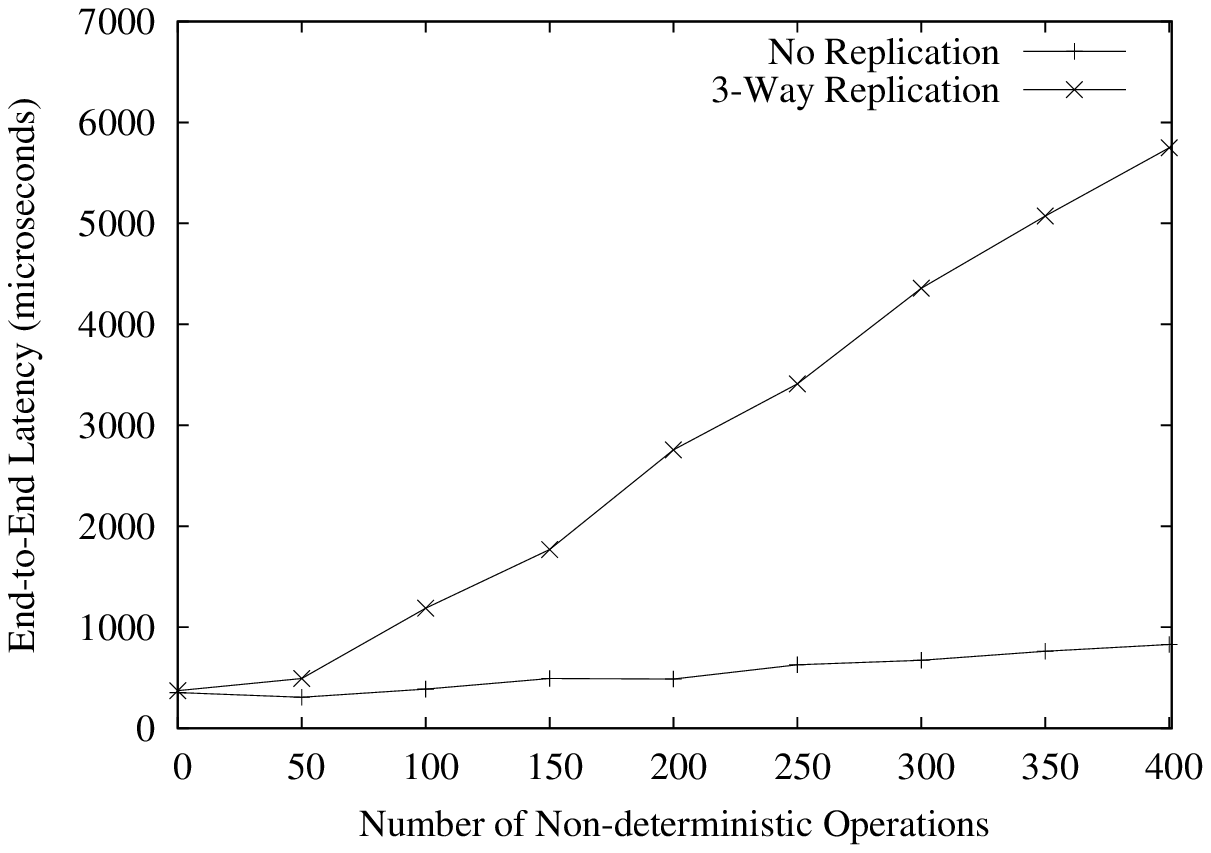}
\vspace{-0.15in} 
\caption{\hspace*{-0.05in}End-to-end latency vs. number of non-deterministic operations.} 
\label{ndlatency}
}}
\vspace{-0.25in}
\end{center}
\end{figure*}

We also characterized the fault scalability of the Messaging
Protocol. As shown in Figure~\ref{scalabilityperf}, the performance
does not degrade noticeably as the number of replicas is increased (so
that larger numbers of concurrent faults can be tolerated). These
results are as expected because the primary can deliver a message as
soon as it is ordered within a connection without having to
communicate with the backups.

\vspace*{-0.1in}
\subsection{Virtual Determinizer Framework}
\vspace*{-0.05in}
To evaluate the performance of the Virtual Determinizer Framework, we
injected non-deterministic operations into our benchmark
application. For each run, we varied the number of non-deterministic
operations per call, while keeping the request/reply message size
fixed at 1KB.

The measurement results for the end-to-end latency shown in
Figure~\ref{ndlatency} are obtained by introducing clock-related
non-deterministic operations (\ie, {\tt gettimeofday()}) into the
application. Other types of non-deterministic operations produce a
similar profile. In general, the end-to-end latency increases linearly
as the number of non-deterministic operations per call increases. On
average, each additional non-deterministic operation adds about 8
microseconds overhead to the end-to-end latency. This
overhead is primarily due to the piggybacking of ordering information.

\vspace*{-0.1in}
\subsection{Membership Protocol}
\vspace*{-0.05in}
To evaluate the performance of the Membership Protocol during recovery
from primary failure, we considered (1) the primary view change
latency, and (2) the recovery latency, {\it i.e.}, the primary view
change latency plus the virtual synchrony latency.  The failover
latency when the primary fails is determined by the fault detection
time and the recovery latency.  In a system that does not incur
lengthy communication delays, the first backup can detect the failure
of the primary in about 30 milliseconds, based on the parameters used
in our experiments.

\begin{figure*}[t]
\begin{center}
\leavevmode
\hbox{\parbox{3.65in}{ 
\epsfxsize=3.6in 
\epsfbox{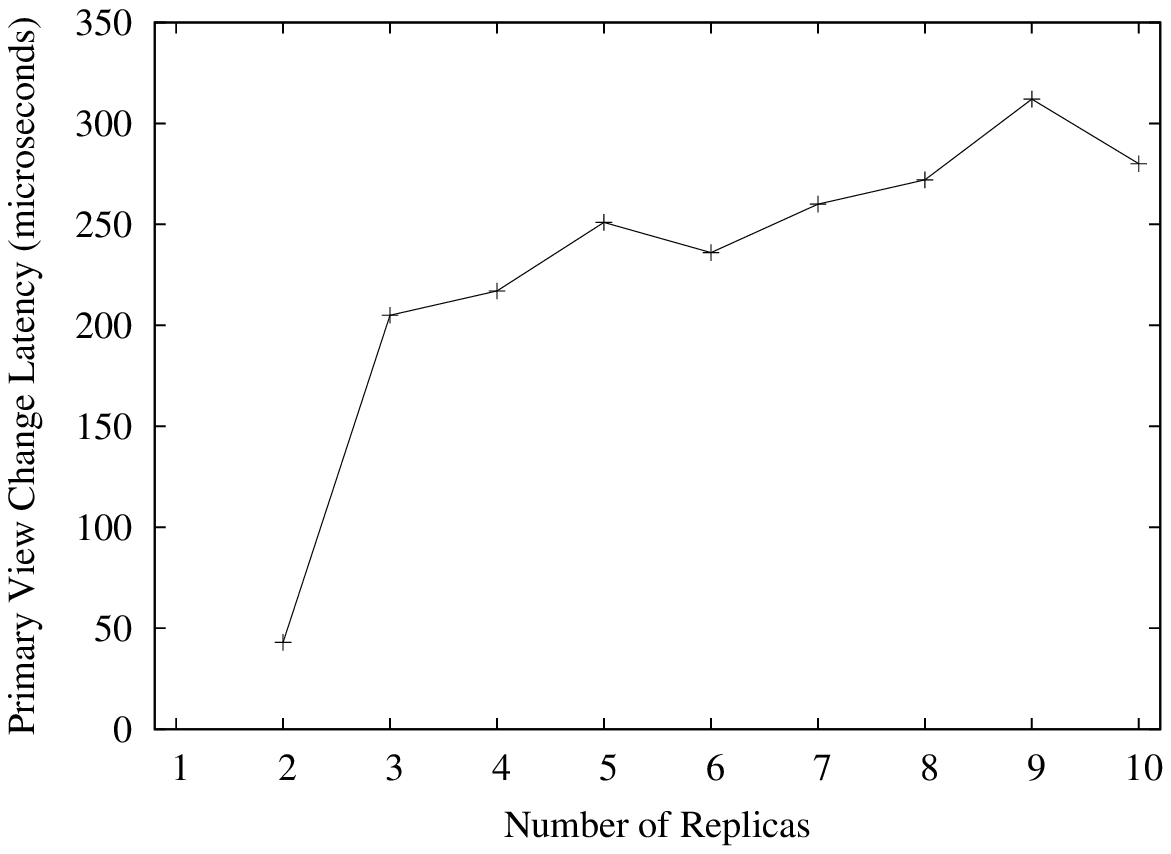}
\vspace{-0.15in}
\caption{Primary view change latency.}
\label{primaryviewchangelatency}
}
\hspace{0.05in}
\parbox{3.65in}{
\epsfxsize=3.6in 
\epsfbox{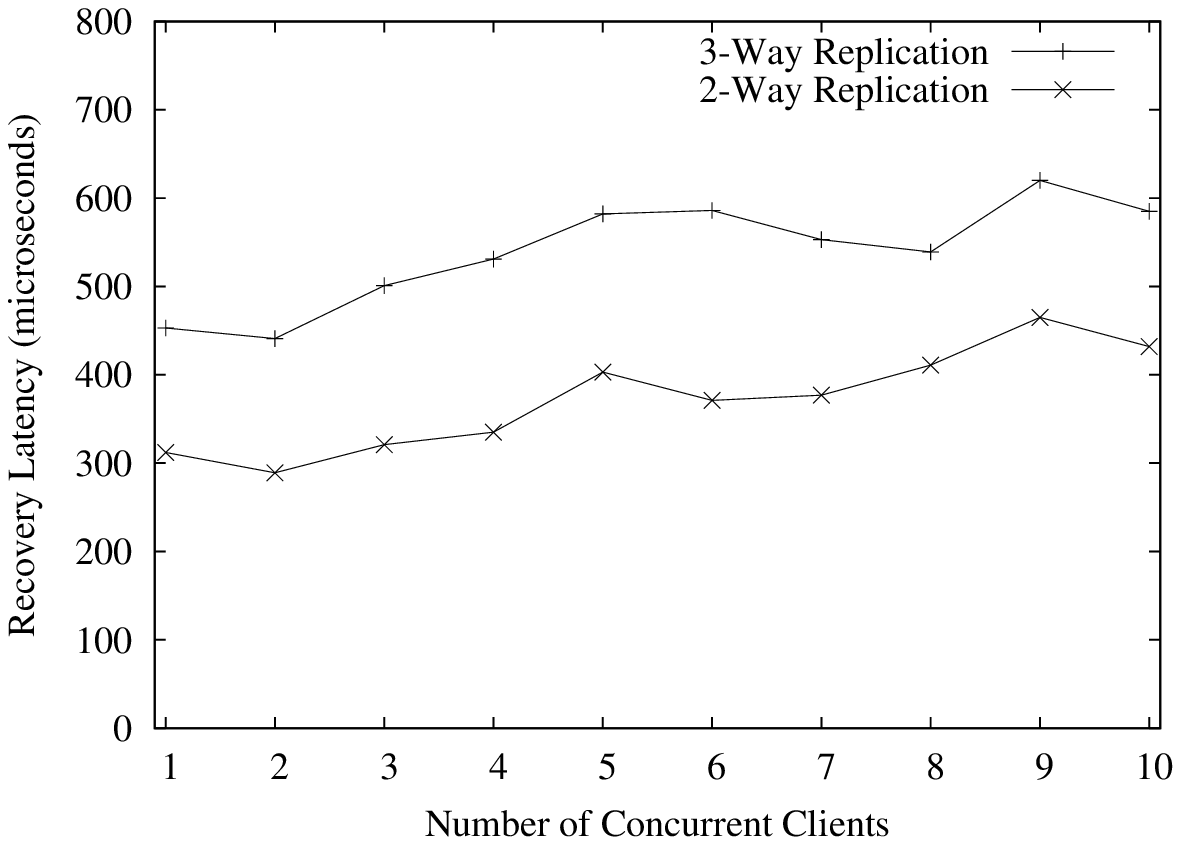}
\vspace{-0.15in}
\caption{Recovery latency.}
\label{recoverylatency}
}}
\vspace{-0.25in}
\end{center}
\end{figure*}

Figure~\ref{primaryviewchangelatency} summarizes the measurement
results for the primary view change latency, which are obtained when
no client is running, to highlight the primary view change latency
itself. As can be seen in the figure, the latency increases with the
number of replicas.  Interestingly, when the number of replicas is two
(which the industry regards as the typical case and which
majority-based membership algorithms do not handle), the primary
view change latency is less than 50 microseconds, which is
significantly less than the latency with more replicas.  In this case,
when the primary crashes, only one replica is left.  That replica can
promote itself to be the new primary without the need to wait for
acknowledgments from other replicas.

Figure~\ref{recoverylatency} summarizes the measurement results for
the recovery latency, {\it i.e.}, the primary view change latency plus
the virtual synchrony latency.  The figure shows the measured recovery
latency in the presence of various numbers of concurrent clients, for
3-way and 2-way replication. As expected, the recovery latency
increases with the number of concurrent clients in both cases.  If the
availability requirement allows 2-way replication (which is typical
industry practice), the recovery is faster by about 200 microseconds.

\vspace*{-0.1in}
\section{Related Work}
\vspace*{-0.05in}
The LLFT system is a software-based approach to fault tolerance.  Such
an approach was first used in SIFT \cite{SIFT} and became the favored
approach in later fault-tolerant systems such as the Delta-4
\cite{Powell:Delta4}, TFT \cite{Bressoud:TFT}, Hypervisor
\cite{hypervisor} and Viewstamped Replication \cite{Liskov} systems.
More recent efforts on software fault tolerance system have focused on
rendering CORBA, Java RMI and Web Services applications fault
tolerant~\cite{birman04,Cukier:AQUA,Felber:Tapos,Baldoni:IRL,tempest-dsn08,moser:ftws,NMM:CSSE,ws-rep,ftweb,WenbingIPDPS2002,zhao:ftws}.
However, supporting a particular messaging protocol API limits the
applicability of those systems.

The LLFT system provides fault tolerance transparently to both the
applications and the operating system, like the TFT
\cite{Bressoud:TFT}, Hypervisor \cite{hypervisor} and TARGON/32
\cite{ftunix} systems. However, those systems differ from LLFT in the
way in which they achieve transparency.  The TARGON/32 system uses a
special bus design that ensures atomic transmission of a message sent
by a primary to both its destination group and its own backups.  The
TFT system requires application object code editing. The Hypervisor
system requires a hardware instruction counter.  LLFT uses the more
flexible library interpositioning technique. The user can dynamically
insert (remove) the LLFT software into (from) a running application
process using operating system facilities.

The LLFT system uses a leader-follower replication technique similar
to that used in the Delta-4 \cite{Powell:Delta4} and Viewstamped
Replication \cite{Liskov} systems.  Delta-4 uses a separate message to
transmit ordering information from the primary to the backups. Thus,
the primary must wait until all of the backups have explicitly
acknowledged the ordering notification before it sends a message,
which can reduce system performance.  In contrast, LLFT uses
piggybacking mechanisms and the Virtual Determinizer Framework to
maintain strong replica consistency and virtual synchrony
\cite{BR:ISIS,Moser}, if a fault occurs.  In the Viewstamped
Replication system, the primary generates a new timestamp each time it
needs to communicate information to the backups.  Unlike LLFT, the
Viewstamped Replication system is based on atomic transactions, which
it combines with a view change algorithm.

Atomic multicast protocols that deliver messages reliably and in the
same total order, such as Isis \cite{BR:ISIS}, Amoeba
\cite{KT:AMOEBA}, and Totem \cite{MMABL:cacm}, have been used to
maintain strong replica consistency in fault-tolerant distributed
systems. However, those protocols introduce delays in either sending
or delivering a message.  The LLFT Messaging Protocol does not incur
such delays, because the primary makes the decisions on the order in
which the operations are performed and the ordering information is
reflected to the backups in its group.  The recent LCR total order
broadcast protocol \cite{Guerraoui2}, which uses logical clocks and a
ring topology, optimizes for high throughput in cluster environments,
rather than low latency as does LLFT.  LCR is comparable to the Totem
single-ring protocol \cite{TotemSRP}, which likewise optimizes for
high throughput in local-area networks, rather than to LLFT.

Paxos \cite{PaxosACMTrans,Paxos} is a leader election algorithm for
asynchronous distributed systems, which uses a two-phase commit
strategy in which a majority of the members must vote for the 
leader.  The requirement of a majority ensures that only one 
leader will be elected.  Paxos assumes a known existing membership,
and does not change that membership dynamically as members become
faulty and recover.  Paxos can achieve consensus in two rounds, if
communication is reliable and processes respond promptly.  There exist
versions of Paxos in which a dedicated proposer selects the new leader
and initiates the election, reducing the number of message delays
required to confirm the new leader, provided that the proposer is not
faulty. There also exist extensions of Paxos in which the leader can
change the membership dynamically, to remove faulty members or to add
new members.  LLFT also employs such ideas.

Membership protocols for group communication systems, such as Transis
\cite{ADKM:MEMB} and Totem \cite{MMABL:cacm} employ fault detectors,
based on timeouts, to reach distributed agreement on as large a
membership as possible, devolving to smaller memberships, if
necessary.  Those protocols are relatively costly in the number of
messages exchanged, and in the delays incurred.  To avoid such costs,
LLFT uses a novel Leader-Determined Membership Protocol that does
involve distributed agreement.  Rather, it achieves consistent group
membership among the members of a group by having the primary
determine the membership, which it communicates to the backups
in the group.

Membership protocols for group communication systems, such as Isis
\cite{BR:ISIS} and Totem \cite{MMABL:cacm}, use the term {\it view
  change} to represent a change in the membership of a group. In
particular, each successive membership, which involves addition
(removal) of a member, constitutes a new view. In LLFT, only
membership changes that correspond to a change of the primary
constitute a new view, which we refer to as a {\it primary view
  change}.  In typical group communication systems, the membership is
known more widely than by only the members in the group.  In LLFT,
only the primary and the backups in the group need to know the
membership of the group.

The LLFT system includes a novel, generic Virtual Determinizer
Framework to capture, transmit and execute ordering information for
non-deterministic operations.  The non-deterministic operations
handled by LLFT overlap those considered in other systems such as
Delta-4 \cite{Powell:Delta4}, TFT \cite{Bressoud:TFT}, Hypervisor
\cite{hypervisor} and TARGON/32 \cite{ftunix}. LLFT addresses
non-determinism caused by multi-threading and socket communication,
which those works do not discuss. LLFT does not yet handle
non-determinism introduced by operating system signals and interrupts,
which those works do consider.

Basile {\it et al.} \cite{iyer:tpds06}, Jimenez-Peris and Arevalo
\cite{Peris:srds}, and Narasimhan {\it et al.} \cite{NMM:SRDS} have
addressed the need to sanitize non-deterministic operations, to
achieve strong replica consistency for active replication, rather than
for leader-follower (semi-active or semi-passive) replication.  The
LLFT mechanisms that are used to order mutex claims/releases are
closely related to those of the Loose Synchronization Algorithm (LSA)
and Preemptive Deterministic Scheduling Algorithm (PDS) of
\cite{iyer:tpds06}.  However, LSA does not address the strong replica
consistency issues introduced by the {\tt pthread\_mutex\_trylock()}
library call, and PDS is suitable for only a specific threading model.

Defago {\it et al.} \cite{Schiper1, Schiper2} have investigated
semi-passive replication in conjunction with a consensus algorithm.
In their model, the primary server produces its results as a single
action, including the reply to the client and the state update for the
backups.  Their model admits non-deterministic operations, but not
concurrent processing, with shared data, of requests from multiple
clients. In our more general model, multiple processes, possibly with
multiple threads, interact with each other and also with file and
database systems.  Moreover, requests from multiple clients can be
processed concurrently and can access shared data.  In their
semi-passive replication, every server replica sends a reply to the
client whereas, in LLFT, the backups do not send replies to the
client, which reduces the amount of network traffic.  Their
semi-passive replication uses a rotating coordinator and a variation
of consensus, whereas LLFT uses a leader-determined membership
protocol that is not based on consensus.

Brito {\it et al.} \cite{FetzerFelberICDCS,FetzerFelberSRDS} have
addressed the issues of minimizing latency and multi-threading in
fault-tolerant distributed stream processing systems.  The system that
they developed supports active replication, instead of the semi-active
or semi-passive replication that LLFT supports.  Their system employs
novel speculation mechanisms based on software transactional memory
to achieve lower latency and higher concurrency.

Zou and Jahanian \cite{Jahanian} have adopted the primary-backup
replication approach for real-time fault-tolerant distributed systems.
Their replication service addresses temporal consistency, introduces
the notion of phase variance, and ensures consistency
deterministically if the underlying communication mechanism provides
deterministic message delivery, and probabilistically if no such
support exists.  LLFT itself provides deterministic message delivery
and ensures strong replica consistency.

\vspace*{-0.1in}
\section{Conclusions and Future Work}
\vspace*{-0.05in} 
The Low Latency Fault Tolerance (LLFT) system provides fault tolerance
for distributed applications deployed over a local-area network, as in
a data center. Applications programmed using TCP socket APIs, or
middleware such as Java RMI, can be replicated with strong replica
consistency using LLFT, without any modifications to the applications.
Performance measurements show that LLFT achieves low latency message
delivery under normal conditions and low latency reconfiguration and
recovery when a fault occurs.  The genericity, application
transparency, and low latency of LLFT make it appropriate for a wide
variety of distributed applications.

Future work includes the sanitization of other sources of
non-determinism (such as operating system signals and
interrupts) and performance optimization.  It also includes the
development of more complex applications for LLFT (in particular, file
systems and database systems), and the development of replication
management tools.

\vspace*{-0.1in}
\appendix
\vspace*{-0.05in}
The proofs of correctness for LLFT, based on the model and the safety
and liveness properties given in Section~\ref{BasicConcepts}, are
provided below.

\newtheorem{theorem}{Theorem}

\begin{theorem}[Safety]
{\it There exists at most one infinite sequence of consecutive
  primary views for the group.  Each of those primary views has a unique primary
  view number and a single primary replica.}
\end{theorem}

\begin{IEEEproof}  
Assume that the primary $R_1$ became faulty in view $V_i$, and that
the backup $R_2$ with precedence $p_2$ and the backup $R_3$ with
precedence $p_3$, where $p_2 < p_3$, each propose a new primary view
with the same primary view number, with itself as the primary of the
new primary view. Consider the following two cases.

Case 1. There is a replica $R$ that is a member of both proposed new
primary views. According to the rules of the LLFT Membership Protocol,
if $R$ first acknowledges $R_3$'s {\tt ProposePrimary} message, then
$R$ does not acknowledge $R_2$'s {\tt ProposePrimary} message, because
$p_2 < p_3$.  Thus, $R_2$ is unable to collect acknowledgments from
all of the proposed members and it abandons its attempt to form that
new membership, resets its state, and applies to rejoin the group.  On
the other hand, if $R$ first acknowledges $R_2$'s {\tt ProposePrimary}
message and subsequently receives $R_3$'s {\tt ProposePrimary}
message, then $R_2$ is unable to collect acknowledgments from all of
the proposed members, because $R_2$ will not receive an acknowledgment
from $R_3$.

Case 2.  There is no replica that is a member of both proposed new
primary views (because neither $R_2$ nor $R_3$ received messages from
any replica in the other's proposed membership and, thus, $R_2$ and
$R_3$ both regard the other's replicas as faulty). Because of the
eventual reliable communication assumption and because of the
retransmission, to all members of the group on the intra-group
connection, of a message containing a higher precedence than the
precedence of the primary of the primary view, every replica $R$ in
$R_2$'s membership eventually receives a message from a replica in
$R_3$'s membership. Because $p_2 < p_3$, $R$ then realizes that
$R_2$'s membership has been superseded by $R_3$'s membership and,
thus, $R$ abandons its current state and applies for readmission to
$R_3$'s membership.  Thus, any side branch is pruned.
\end{IEEEproof}  

In the theorems and proofs below, operations refer to both computation
and communication operations.

\begin{theorem}[Safety]
{\it There exists at most one infinite sequence of operations
  in an infinite sequence of consecutive primary views for the group.}
\end{theorem}

\begin{IEEEproof} 
By Theorem 1, there exists at most one infinite sequence of
consecutive primary views.  Each of those primary views has a unique
primary view number and a single primary replica. Moreover, each of
those primary view has an associated sequence of operations determined
by the primary of that view.  The sequence of operations in an
infinite sequence of consecutive primary views is the concatenation of
the sequences of operations for the primary views in the order of
their primary view numbers.
\end{IEEEproof} 

\newtheorem{lemma}{Lemma}

\begin{lemma}
{\it For semi-active replication, if a backup replica $R_2$ is
  admitted to a membership of view $V_i$ by primary replica $R_1$ after
  the start of $V_i$ and $R_2$ subsequently becomes faulty in
  $V_i$, then the sequence of operations of $R_2$ in $V_i$ is a
  consecutive subsequence of the sequence of operations of $R_1$ in
  $V_i$.}
\end{lemma}

\begin{IEEEproof} 
For primary replica $R_1$ of view $V_i$, the sequence of ordering
information is determined by the sequence of operations of $R_1$.
When the backup $R_2$ is admitted to a membership of $V_i$ by $R_1$
after the start of view $V_i$, it receives a {\tt State} message from
$R_1$, which establishes a synchronization point. After that point,
the sequence of operations performed by $R_2$ is determined by the
sequence of ordering information provided by $R_1$, until $R_2$
becomes faulty.  Thus, the sequence of operations of $R_2$ in $V_i$ is
a consecutive subsequence of the sequence of operations of $R_1$ in
$V_i$.
\end{IEEEproof} 

\begin{lemma}
{\it For semi-active replication, if replica $R_2$ is admitted to the
  membership of view $V_i$ by primary replica $R_1$ after the start of
  $V_i$ and $R_2$ is not faulty in $V_i$, then the sequence of operations of
  $R_2$ in $V_i$ is a suffix of the sequence of operations of $R_1$ in
  $V_i$.}
\end{lemma}

\begin{IEEEproof} 
For primary replica $R_1$ of view $V_i$, the sequence of ordering
information is determined by the sequence of operations of $R_1$.
When $R_2$ is admitted to the membership of view $V_i$ by $R_1$ after
the start of $V_i$, it receives a {\tt State} message from $R_1$,
which establishes a synchronization point. After that point, the
sequence of operations of $R_2$ is determined by the sequence of
ordering information provided by $R_1$ in $V_i$.  Moreover, because
$R_2$ is not faulty in $V_i$, it participates in the virtual synchrony
at the start of $V_{i+1}$.  Thus, $R_2$'s sequence of operations in
$V_i$ is a suffix of the sequence of operations of $R_1$ in $V_i$.
\end{IEEEproof} 

\begin{lemma}
{\it For semi-active replication, if replica $R_2$ is an initial
  member of view $V_i$ with primary replica $R_1$ and $R_2$
  subsequently becomes faulty in $V_i$, then the sequence of
  operations of $R_2$ in $V_i$ is a prefix of the sequence of
  operations of $R_1$ in $V_i$.}
\end{lemma}

\begin{IEEEproof} 
For primary replica $R_1$ of view $V_i$, the sequence of ordering
information is determined by the sequence of operations of $R_1$.
Because $R_2$ is an initial member of view $V_i$, it participates in
the virtual synchrony at the start of $V_i$.  After that point, the
sequence of operations of $R_2$ is determined by the sequence of
ordering information provided by $R_1$ in $V_i$, until $R_2$ becomes
faulty.  Thus, $R_2$'s sequence of operations in $V_i$ is a prefix of
the sequence of operations of $R_1$ in $V_i$.
\end{IEEEproof} 

\begin{lemma}
{\it For semi-active replication, if replicas $R_2$ and $R_3$ are
  members of the same memberships of views $V_i$, $V_{i+1}$ and
  $V_{i+2}$, then the sequence of operations of $R_2$ in $V_{i+1}$ is
  the same as the sequence of operations of $R_3$ in $V_{i+1}$.}
\end{lemma}

\begin{IEEEproof} 
Because $R_2$ and $R_3$ are members of the same memberships of views
$V_i$ and $V_{i+1}$, both of them participate in the virtual synchrony
between $V_i$ and $V_{i+1}$, determined by primary $R_1$ of $V_{i+1}$.
Because both $R_2$ and $R_3$ are members of the same memberships of
views $V_{i+1}$ and $V_{i+2}$, neither of them becomes faulty in
$V_{i+1}$ and both of them participate in the virtual synchrony
between $V_{i+1}$ and $V_{i+2}$.  Consequently, both $R_2$ and $R_3$
perform the same sequence of operations in $V_{i+1}$, which is the
same as the sequence of operations performed by $R_1$, determined by
the sequence of ordering information provided by $R_1$.
\end{IEEEproof} 

\begin{lemma}
{\it For semi-active replication, if replicas $R_2$ and $R_3$ are
  members of the same memberships of views $V_i$ and $V_k$, then the
  sequence of operations of $R_2$ in $V_j$ is the same as the sequence
  of operations of $R_3$ in $V_j$, where $i < j < k$.}
\end{lemma}

\begin{IEEEproof} 
Because $R_2$ and $R_3$ are members of the same membeships of views
$V_i$ and $V_k$, they are both members of the same memberships of
views $V_j$ for all $j$, $i < j < k$, because if they are removed from
a membership and apply for readmission, they are admitted as new
replicas.  The proof now follows from Lemma 4 by induction.
\end{IEEEproof} 

\begin{theorem}[Safety] 
{\it For semi-active replication, the sequence of operations of a
  replica, in a membership of the infinite sequence of consecutive
  primary views, is a consecutive subsequence of the infinite sequence
  of operations for the group.}
\end{theorem}

\begin{IEEEproof} 
We consider the following three cases for a semi-active replica $R$.

Case 1: Replica $R$ is admitted to a membership in view $V_i$ and
becomes faulty in the same view $V_i$.  By Lemma 1, the sequence of
operations of $R$ in $V_i$ is a consecutive subsequence of the
sequence of operations of primary replica $R_1$ in $V_i$ and, thus, of
the infinite sequence of operations for the group.

Case 2: Replica $R$ is admitted to a membership in view $V_i$ and
becomes faulty in view $V_{i+1}$.  By Lemma 2, the sequence of
operations of $R$ in $V_i$ is a suffix of the sequence of operations
of primary replica $R_1$ in $V_i$.  By Lemma 3, the sequence of
operations of $R$ in $V_{i+1}$ is a prefix of the sequence of
operations of primary replica $R_1^{\prime}$ in $V_{i+1}$.  Thus, the
sequence of operations of $R$ is the concatenation of the suffix for
view $V_i$ and the prefix for view $V_{i+1}$. Thus, the sequence of
operations of $R$ is a consecutive subsequence of the infinite
sequence of operations for the group.

Case 3: Replica $R$ is admitted to the membership in view $V_i$ and
becomes faulty in view $V_k$, where $k > i+1$.  By Lemma 2, the
sequence of operations of $R$ in $V_i$ is a suffix of the sequence of
operations of primary replica $R_1$ in $V_i$.  By Lemma 3, the sequence of
operations of $R$ in $V_k$ is a prefix of the sequence of operations
of primary replica $R_1^{\prime}$ in $V_k$.  By Lemma 5, the sequence of
operations of $R$ in $V_j$ is the same as the sequence of operations
of primary replica $R_1^{{\prime}{\prime}}$ in $V_j$, where $i < j < k$.
Thus, the sequence of operations of $R$ is the concatenation of the
suffix for view $V_i$, the sequences for the views $V_j$, where $i < j
< k$, and the prefix for view $V_k$. Thus, the sequence of operations
of $R$ is a consecutive subsequence of the infinite sequence of
operations for the group.
\end{IEEEproof} 

In the proofs above, which apply to semi-active replication, we
consider the sequence of operations performed at the primary and at
the backups.  To address semi-passive replication, we consider the
sequence of states at the primary and the backups, because for
semi-passive replication, the backups perform no operations.

\begin{theorem}[Safety]
{\it For semi-passive replication, the sequence of states of a
  replica, in a membership of the infinite sequence of consecutive
  primary views, is a consecutive subsequence of the infinite sequence
  of states for the group.}
\end{theorem}

\begin{IEEEproof} 
The proof is similar to that for Theorem 3 for semi-active
replication.
\end{IEEEproof} 

\begin{theorem}[Liveness]
{\it There exists at least one infinite sequence of consecutive
  primary views with consecutive primary view numbers for the group.}
\end{theorem}

\begin{IEEEproof} 
By the sufficient replication assumption ({\it i.e.}, each
group contains enough replicas such that in each primary view there
exists at least one replica that does not become faulty), if the primary becomes
faulty in a view $V_i$, then there exists a replica $R$ in $V_i$ that
can assume the role of the primary in view $V_{i+1}$.  The proof now
follows by induction.
\end{IEEEproof} 

\begin{theorem}[Liveness]
{\it There exists at least one infinite sequence of operations in each
  infinite sequence of consecutive primary views for the group.}
\end{theorem}

\begin{IEEEproof} 
There exists at least one operation (the communication of the {\tt
  State} message) in each primary view.  The proof now follows from
Theorem 5.
\end{IEEEproof} 

\vspace*{-0.1in}
\section*{Acknowledgment}
\vspace*{-0.05in}
This research was supported in part by NSF grant CNS-0821319, and by
a CSISI grant from Cleveland State University (for the first author).

\bibliographystyle{IEEETran}
\renewcommand{\baselinestretch}{1.0}

\end{document}